\documentclass[reprint, amsmath, amssymb, prx, superscriptaddress, floatfix]{revtex4-2}

\usepackage{upgreek}
\usepackage{subfigure}

\usepackage[per-mode=symbol,separate-uncertainty]{siunitx}
\usepackage[english]{babel}
\usepackage[pdftex]{graphicx} 
\usepackage[]{graphics}
\usepackage{amsmath,color,colordvi}  
\usepackage{mathtools}  
\usepackage{braket}
\usepackage{textcomp}
 
\definecolor{navyblue}{rgb}{0.0, 0.0, 0.5}
\usepackage[protrusion=true, expansion=true]{microtype} 
\usepackage[colorlinks=true, breaklinks=false, linkcolor=navyblue, urlcolor=navyblue, citecolor=navyblue]{hyperref}

\begin{document}

\renewcommand{\figurename}{FIG.}
\renewcommand{\tablename}{TABLE}

\title{Photon-number-dependent effective Lamb shift}

\author{Arto Viitanen}
\email[To whom correspondence should be addressed; E-mail: arto.viitanen@aalto.fi and mikko.mottonen@aalto.fi.]{}
\affiliation{QCD Labs, QTF Center of Excellence, Department of Applied Physics, Aalto University, P.O. Box 13500, FI-00076 Aalto, Finland.}
\author{Matti Silveri}
\affiliation{Research Unit of Nano and Molecular Systems, University of Oulu, P.O. Box 3000, FI-90014 Oulu, Finland.}
\author{M\'at\'e Jenei}
\affiliation{QCD Labs, QTF Center of Excellence, Department of Applied Physics, Aalto University, P.O. Box 13500, FI-00076 Aalto, Finland.}
\affiliation{IQM Finland Oy, Keilaranta 19, FI-02150 Espoo, Finland.}
\author{Vasilii Sevriuk}
\affiliation{QCD Labs, QTF Center of Excellence, Department of Applied Physics, Aalto University, P.O. Box 13500, FI-00076 Aalto, Finland.}
\affiliation{IQM Finland Oy, Keilaranta 19, FI-02150 Espoo, Finland.}
\author{Kuan Y. Tan}
\affiliation{QCD Labs, QTF Center of Excellence, Department of Applied Physics, Aalto University, P.O. Box 13500, FI-00076 Aalto, Finland.}
\affiliation{IQM Finland Oy, Keilaranta 19, FI-02150 Espoo, Finland.}
\author{Matti~Partanen}
\affiliation{QCD Labs, QTF Center of Excellence, Department of Applied Physics, Aalto University, P.O. Box 13500, FI-00076 Aalto, Finland.}
\author{Jan Goetz}
\affiliation{QCD Labs, QTF Center of Excellence, Department of Applied Physics, Aalto University, P.O. Box 13500, FI-00076 Aalto, Finland.}
\affiliation{IQM Finland Oy, Keilaranta 19, FI-02150 Espoo, Finland.}
\author{Leif Gr\"onberg}
\affiliation{VTT Technical Research Centre of Finland Ltd., QTF Center of Excellence, P.O. Box 1000, FI-02044 VTT, Finland.}
\author{Vasilii Vadimov}
\affiliation{QCD Labs, QTF Center of Excellence, Department of Applied Physics, Aalto University, P.O. Box 13500, FI-00076 Aalto, Finland.}
\affiliation{MSP Group, QTF Centre of Excellence, Department of Applied Physics, Aalto University, P.O. Box 11000, FI-00076 Aalto, Finland}
\affiliation{Institute for Physics of Microstructures, Russian
Academy of Sciences, 603950 Nizhny Novgorod, GSP-105, Russia}
\author{Valtteri Lahtinen}
\affiliation{QCD Labs, QTF Center of Excellence, Department of Applied Physics, Aalto University, P.O. Box 13500, FI-00076 Aalto, Finland.}
\author{Mikko M\"{o}tt\"{o}nen}
\email[To whom correspondence should be addressed; E-mail: arto.viitanen@aalto.fi and mikko.mottonen@aalto.fi.]{}
\affiliation{QCD Labs, QTF Center of Excellence, Department of Applied Physics, Aalto University, P.O. Box 13500, FI-00076 Aalto, Finland.}
\affiliation{VTT Technical Research Centre of Finland Ltd., QTF Center of Excellence, P.O. Box 1000, FI-02044 VTT, Finland.}

\date{\today}

\begin{abstract} 
The Lamb shift, an energy shift arising from the presence of the electromagnetic vacuum, has been observed in various quantum systems and established as the part of the energy shift independent of the environmental photon number. However, typical studies are based on simplistic bosonic models which may be challenged in practical quantum devices. We demonstrate a hybrid bosonic-fermionic environment for a linear resonator mode and observe that the photon number in the environment can dramatically increase both the dissipation and the effective Lamb shift of the mode. Our observations are quantitatively described by a first-principles model which we develop here also to guide device design for future quantum-technological applications. The device demonstrated here can be utilized as a fully rf-operated quantum-circuit refrigerator to quickly reset superconducting qubits.
\end{abstract}

\maketitle

\section{Introduction}

Quantum theory, put forward a century ago has not only revealed the deepest secrets of nature such as those of the elementary particles but also given rise to a multitude of practical applications ranging from medical imaging~\cite{Ernst1987,Hamalainen1993} to the emerging quantum computers~\cite{Ladd10,Neill18,Rosenblum18,Arute2019,Blais20} and quantum cryptography~\cite{Bennett84,Ekert91}. One of the most important challenges in contemporary quantum physics is the understanding and control of the effects of environments on the studied quantum systems.

In 1947, Willis Lamb and Robert Retherford conducted their celebrated experiments to observe an anomalous small shift in the energy spectrum of hydrogen~\cite{Lamb47}. This frequency shift, later referred to as the Lamb shift, was considered to arise from the sole presence of the electromagnetic vacuum, not from excitations of the field~\cite{Bethe47}. It paved the way for the golden ages of physics where modern quantum electrodynamics and particle physics were developed along with discoveries such as a theory of superconductivity~\cite{Bardeen57}, the Casimir effect~\cite{Casimir48}, and applications of quantum tunnelling~\cite{Giaever60}. Yet, the Lamb shift still ties the contemporary quantum physicists and engineers with the importance of the quantum environment: quantum systems are not isolated---even if energy relaxation can be mitigated by filtering at the system frequency, the Lamb shift arising from the full spectrum of the environmental modes persists.

In fact, the Lamb shift has not only been studied in natural quantum systems such as atoms~\cite{Lamb47,Bethe47,Heinzen87,Brune94,Marrocco98,Rentrop16}, but has also fascinated physicists working on engineered quantum systems. The Lamb shift in superconducting qubits was first observed as a result of a single bosonic mode~\cite{Fragner08, Yoshihara18} and later arising from a few well-defined modes~\cite{Mirhosseini18}. Also the Lamb shift owing to lattice vibrations, or phonons, has been observed in ultracold quantum gases~\cite{Rentrop16}. Recently, the studies of the Lamb shift in microwave circuits were taken a step closer to Lamb's original experiments, namely, to broadband environments~\cite{Silveri19}. This was achieved with a quantum-circuit refrigerator~\cite{Tan16} (QCR) which was used as a tunable environment consisting of a continuum of modes.

The Lamb shift originates from the coupling between the environmental modes and the studied system, even though the modes are not excited~\cite{Lamb47,Bethe47,Carmichael99}. In contrast, the shift that stems from the occupation of the modes is generally referred to as the ac Stark shift~\cite{Carmichael99}. If a transition frequency of the studied system depends non-linearly on a coupled external field, fluctuations in the field give rise to a different average frequency compared with the uncoupled system. Thus the higher the occupation in the environmental modes, the more the field it induces effectively fluctuates, and the more ac Stark shift is observed. Importantly, the ac Stark shift is absent in a linear system linearly coupled to its environment~\cite{Carmichael99}. Thus such a linear system is ideal for studying the Lamb shift which persists.

\begin{figure*}
    \centering
	\includegraphics[width=2\columnwidth]{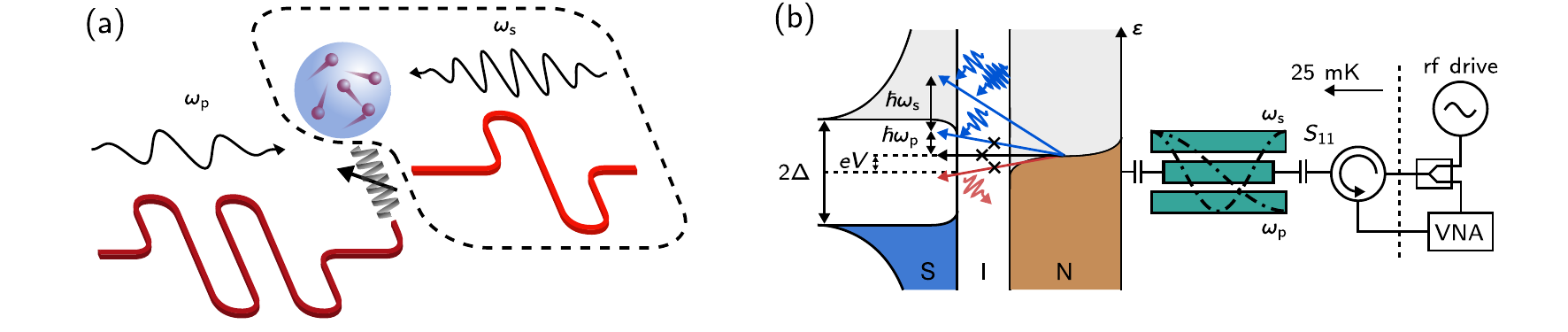}
	\caption{\label{fig:setup}Studied system and measurement scheme. (a) Illustration of the interaction between the primary bosonic mode at angular frequency $\omega_{\rm p}$ and its engineered environment (dashed box) consisting of a quantum-circuit refrigerator (QCR) and a supporting mode at $\omega_{\rm s}$. (b) Schematics of a coplanar waveguide resonator (green) coupled to the normal-metal island of the QCR on the left and to the measurement setup on the right. The dash-dotted lines show the voltage amplitudes of the primary and supporting modes. A vector network analyzer (VNA) measures the reflection coefficient $S_{11}$ (see Appendix~\ref{sec:measurement}). The energy diagram of the QCR illustrates photon-assisted electron tunnelling at a normal-metal--insulator--superconductor (NIS) junction. The electron occupation in the normal-metal (brown shading) follows the Fermi distribution, whereas the superconductor states are essentially filled (blue shading) up to the Bardeen--Cooper--Schrieffer energy gap of $2\Delta$. States at high energies $\varepsilon$ are vacant. Photon-assisted tunnelling events can absorb photons from (blue arrows) or emit to (red arrow) the primary mode, of energy $\hbar\omega_{\rm p}$, or from the supporting mode, of energy $\hbar\omega_{\rm s}$. Elastic events (black arrow) do not directly induce dissipation on the resonator. Energetically forbidden processes at the example dc bias voltage, denoted by $V$, are crossed out.}
\end{figure*}

Typically, as observed in the case of linear bosonic environments~\cite{Lamb47,Bethe47,Heinzen87,Brune94,Marrocco98,Rentrop16,Silveri19,Wen2019}, the Lamb shift depends on the coupling strength between the system and the environment and on the density of the environmental states, but not on the photon number or the temperature of the environment. In this paper, however, we implement a hybrid~\cite{Clerk20,Lachance20,Barzanjeh19} environment with bosonic and fermionic constituents. Namely, we couple our studied system, which is a mode of a superconducting resonator, to the environment consisting of a QCR and another mode of the resonator (Figs.~\ref{fig:setup} and \ref{fig:circuit}). The system mode is referred to as the primary mode and the environmental mode as the supporting mode. We measure the resonance frequency of the primary mode at different QCR bias voltages and average photon numbers of the supporting mode (Fig.~\ref{fig:F_landscape}). We observe that the frequency strongly depends on not only the bias voltage~\cite{Silveri19} but also on the photon number (Fig.~\ref{fig:results}). This observation of a photon-number-dependent frequency shift apparently contradicts the typical non-hybrid case highlighting the observed novel physics owing to the hybrid nature of the environment: an increase in the photon number in the bosonic part of the environment effectively manifests in the primary mode as an increased environmental coupling strength mediated by the fermionic part, and consequently as an increased effective Lamb shift. Finally, we quantitatively explain using first-principles quantum theory (see Appendices~\ref{sec:rfqcr} and~\ref{sec:lamb}) the origin of a novel oscillatory behavior of the effective Lamb shift as a function of the bias voltage. 

\section{Results}
\subsection{Radio frequency quantum-circuit refrigerator}

\begin{figure}
    \centering
	\includegraphics[width=\columnwidth]{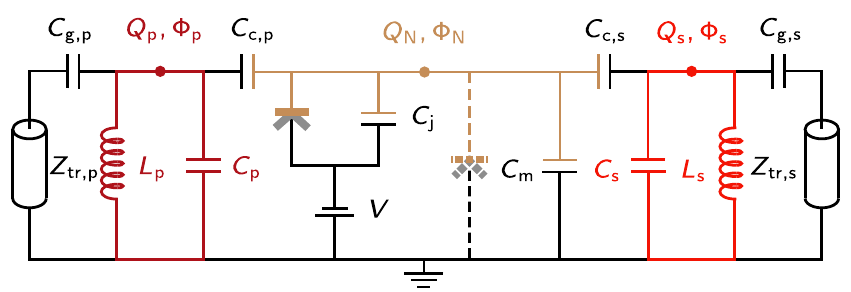}
	\caption{\label{fig:circuit}Lumped-element circuit model of the studied system. The primary and the supporting mode are each modeled by an $LC$ resonator (dark and bright red) with capacitances $C_{\rm p}$ and $C_{\rm s}$, and inductances $L_{\rm p}$ and $L_{\rm s}$, respectively. The resonators are coupled to the normal-metal island of the QCR (brown color) through capacitances $C_{\rm c,p} = C_{\rm c,s}$. Tunnelling occurs between the normal metal and the superconducting electrode (grey color). This junction is associated with the capacitance $C_{\rm j}$ and biased with $V$, allowing voltage control over the photon-assisted electron tunnelling. The effects of the other parallel NIS junction (dashed lines) are included in the capacitance $C_{\rm m}$ of the normal-metal island to the ground and in the value of the junction resistance used in the model. In the experimental sample, we have a single physical resonator, which implies that in this lumped-element model we have to use equal output capacitances $C_{\rm g,p} = C_{\rm g,s}$ and characteristic impedances $Z_{\rm tr,p} = Z_{\rm tr,s}$ to correctly model the external coupling of the considered resonator modes. The classical node fluxes at the island and at the resonators are denoted by $\Phi_{\rm N},\Phi_{\rm p},$ and $\Phi_{\rm s}$, and their conjugate charges by $Q_{\rm N}, Q_{\rm p},$ and $Q_{\rm s}$, respectively.}
\end{figure}

Our physical system is described in Figures~\ref{fig:setup} and \ref{fig:circuit} (see also Fig.~\ref{fig:sampleimage} in Appendix~\ref{sec:fabrication}) with parameters listed in Table~\ref{tb:parameters}. The sample consists of a superconducting coplanar waveguide (CPW) resonator capacitively coupled to a QCR, i.e., to a normal-metal island that forms tunnel junctions with two superconducting leads. In general, a QCR provides tunable dissipation for the electromagnetic excitations of the coupled quantum circuit which is in our case the CPW resonator with two considered modes. The dissipation is controlled with a bias voltage applied through the two leads equipped with on-chip low-pass \textit{RC} filters in order to keep the dc line environment well controlled. The fabrication of the sample is discussed in Appendix~\ref{sec:fabrication}.

\begin{table}
	\centering
    \caption{\label{tb:parameters}Key device and model parameters. See Appendix~\ref{sec:analysis} for details of the experimental determination of the parameters.}
    \begin{tabular}{l|c|c|c} 
    \hline
    Parameter & Symbol & Value & Unit \\  
    \hline  
    Resonator frequency & $\omega_{\rm p}/2\pi$ & $8.8241$  & \si{\giga\hertz}\\
    Supporting mode frequency & $\omega_{\rm s}/2\pi$ & $17.651$  & \si{\giga\hertz} \\
    Resonator characteristic impedance & $Z_{\rm p}$ & $42.8$ & \si{\ohm} \\
    External coupling strength & $\gamma_{\rm tr,p}/2\pi$ & $2.1$ & \si{\mega\hertz} \\ 
    Excess coupling strength & $\gamma_{\rm 0,p}/2\pi$ &  $1.6$ & \si{\mega\hertz} \\
    Coupling capacitance & $C_{\rm c,p}$ & $780$  & \si{\femto\farad} \\
    Output capacitance & $C_{\rm g,p}$ & $6.4$ & \si{\femto\farad} \\
    Island capacitance & $C_{\rm \Sigma}$ & $4$  & \si{\femto\farad} \\
    Primary-mode capacitance ratio & $\alpha_{\rm p}$ & $0.7817$  \\
    Supporting-mode capacitance ratio & $\alpha_{\rm s}$ & $0.7413$  \\
    Superconductor gap parameter & $\Delta $ & $208$ & \si{\micro\eV} \\
    Dynes parameter & $\gamma_{\rm D}$ & \num[output-product=\times]{4xe-4}  \\
    Electron temperature & $T_{\rm N}$ & $90$ & \si{\milli\kelvin} \\
    \hline 
  \end{tabular}
\end{table}

As depicted in Fig.~\ref{fig:setup}(b), the interaction between the fundamental mode of the resonator (resonance frequency $\omega_{\rm p}/2\pi = 8.8241$~\si{\giga\hertz} $\pm~0.18$~\si{\mega\hertz}, linewidth $1.9$~MHz) and its engineered environment is mediated by photon-assisted electron-tunnelling events through the normal-metal--insulator--superconductor (NIS) junctions in the QCR. Instead of a single NIS junction, we employ a pair of junctions to double the tunnelling rate as well as to supply a return path to the bias current from the normal-metal island. The tunnelling processes are activated if the electrons are supplied with sufficient additional energy to compensate for the superconductor energy gap of $2\Delta$ around the Fermi level. The energy can be provided continuously with the tunable bias voltage, yielding $eV$, and in a quantized manner by $n_{\rm p}$ absorbed photons from the fundamental mode, yielding $n_{\rm p}\hbar\omega_{\rm p}$. With the inclusion of the rf-driven second mode ($\omega_{\rm s}/2\pi = 17.651$~\si{\giga\hertz}, linewidth $11$~MHz), slightly shifted from $2\times\omega_{\rm p}$ due to the coupling capacitance, the bias voltage is no longer strictly necessary to activate tunnelling events even for vanishing $n_{\rm p}\hbar\omega_{\rm p}$. Instead, the tunnelling electrons may acquire the necessary energy, in excess of $n_{\rm p}\hbar\omega_{\rm p}$, from the photons of the second bosonic mode. Hence, we refer to the fundamental mode as the primary mode (subscripts $\rm{p}$) and to the second mode as the supporting mode (subscripts $\rm{s}$).

An electron-tunnelling event at the junctions induces a charge shift of $\Delta Q=\alpha_{\rm p/s}e$ on the resonator modes. The strength of the effect is dependent on the circuit parameters (see Fig.~\ref{fig:circuit}) and determined by the constant $0<\alpha_{\rm p/s}<1$, which is related to the capacitances of the resonator modes, $C_{\rm p/s}$, and of the couplings to the QCR, $C_{\rm c,p/s}$ (see Appendix~\ref{sec:rfqcr}). The charge shift couples the different eigenstates of the two modes and may induce transitions between them, resulting in photon-assisted electron tunnelling with annihilation or creation of photons in the two modes, as illustrated in Fig.~\ref{fig:setup}(b). 
 
\begin{figure*}
	\centering
	\includegraphics[width=2\columnwidth]{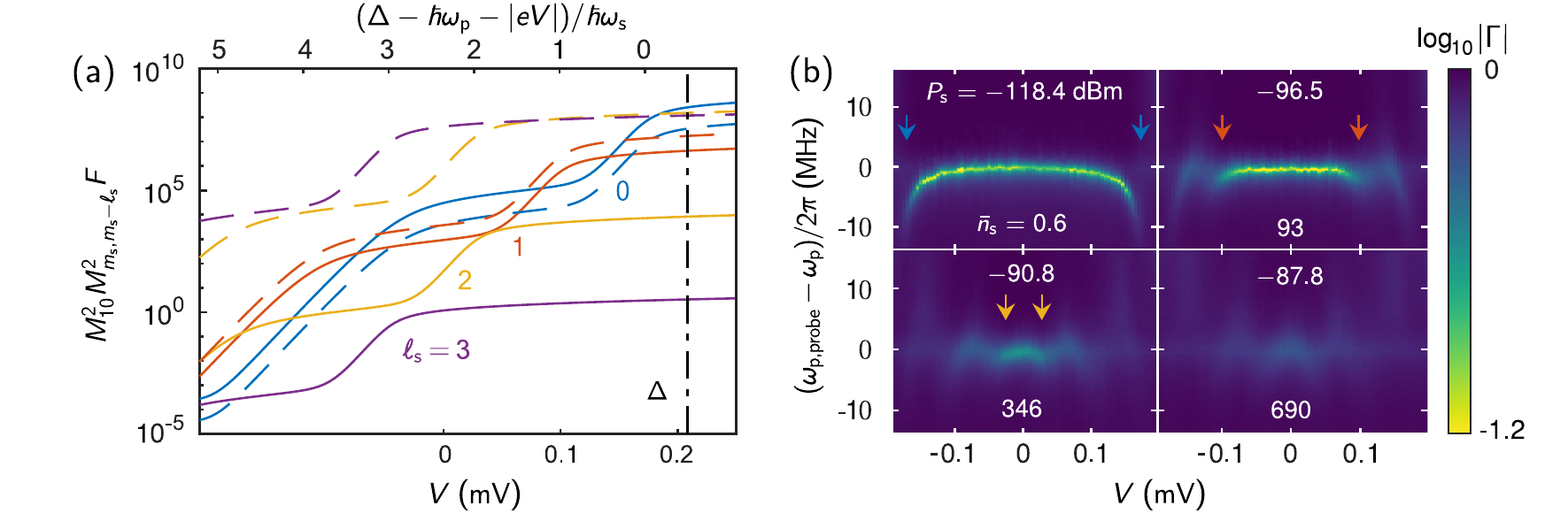}
	\caption{\label{fig:F_landscape}Onset of multiphoton processes.	(a) Relaxation rate of the primary mode from the state $\ket{1}$ to $\ket{0}$ through processes involving $\ell_{\rm s} = 0,1,2,3$ supporting-mode photons as a function of the bias voltage for small (solid lines, $m_{\rm s}=3$) and large (dashed lines, $m_{\rm s}=1000$) initial supporting-mode occupation $m_{\rm s}$. Top axis shows the energy needed from the supporting-mode photons to activate the QCR for the primary mode. (b) Reflection magnitude as a function of the bias voltage and probe frequency from negligible supporting-mode drive strength (top left panel) to strong drive (bottom right panel) with the drive power levels at the sample input and the estimated supporting-mode steady-state average photon number $\bar{n}_{\rm s}(V=0)$ indicated. The reflection magnitude is normalized such that its maximum value is one in each panel. The background has been corrected as described in Appendix~\ref{sec:analysis}. The arrows indicate the bias points, $V=(\Delta-\hbar\omega_\textrm{p}-\ell_\textrm{s}\hbar\omega_\textrm{s})/e$, where relaxation processes involving $\ell_\textrm{s}=0$ (blue), 1 (red), and 2 (orange) supporting-mode photons become active.}
\end{figure*}

For the sake of simplicity, we assume that the two NIS junctions are identical, the corresponding electrodes are at equal temperatures, and the charging energy of the normal-metal island, $E_\textrm{N}$, is small compared to the other relevant energy scales. After tracing out the supporting mode, the electromagnetic environment of the primary mode is  characterized by its effective coupling strength to the hybrid environment (see Appendix~\ref{sec:rfqcr})
\begin{widetext}
\begin{equation}
	\label{eq:gamma_main}
	\gamma_{\rm T,p}(V,\bar{n}_{\rm s}) = 2\pi\alpha_{\rm p}^2\frac{Z_{\rm p}}{R_{\rm{T}}}\sum_{k,l}P_k(\bar{n}_{\rm s})\left|M_{kl}^{(\textrm{s})}\right|^2\sum_{\ell_{\rm p},\tau=\pm1}\ell_{\rm p} F(\tau eV+\ell_{\rm p}\hbar\omega_{\rm p}+\ell_{\rm s}\hbar\omega_{\rm s}-E_{\rm N}), 
\end{equation}
\end{widetext}
where $Z_{\rm p}$ is the characteristic impedance of the primary mode, $R_{\rm T}$ is the tunnelling resistance, $P_k$ is the occupation probability of the $\it{k}$th supporting-mode eigenstate, $\bar{n}_{\rm s}=\sum_k kP_k$ is the mean supporting-mode photon number, $M_{kl}^{({\rm s})}$ is the transition matrix element from the $\it{k}$th to the $\it{l}$th supporting-mode eigenstate, $\ell_{\rm s} = k-l$ is the number of supporting-mode photons absorbed by the electron-tunnelling event, $F$ is the normalized forward tunnelling rate (see Appendix~\ref{sec:rfqcr}), and $\tau$ takes into account both direction of the electron tunneling through the NIS junction. We excite the supporting mode with a classical drive, yielding $P_k(\bar{n}_{\rm s}) = \textrm{e}^{-\bar{n}_{\rm s}}\frac{\bar{n}_{\rm s}^k}{k!}$.

\subsection{Multiphoton-assisted electron tunnelling}

The argument of the normalized forward tunnelling rate $F$ in equation~\eqref{eq:gamma_main} is the energy obtained in the tunnelling process by the electrons from the resonator modes and from the voltage source. The remaining energy for tunnelling is provided by thermal excitations, described by the Fermi functions in the superconducting electrodes and the normal-metal island (see Appendix~\ref{sec:rfqcr}). Figure~\ref{fig:F_landscape}(a) shows three characteristic regions~\cite{Silveri17} for $F(\delta E)$: (i) For $\delta E\ll\Delta$, the electron tunnelling and the resulting coupling strength $\gamma_{\rm T,p}$ are suppressed by the gap in the density of states of the superconductor. (ii) Once $\delta E$ is raised sufficiently near the gap edge, the photon-assisted electron tunnelling is thermally activated, and the coupling strength increases exponentially, (iii) before saturating near the gap edge.

The driven supporting mode promotes multiphoton-assisted electron-tunnelling events, as illustrated in Fig.~\ref{fig:setup}(b) and quantified in Fig.~\ref{fig:F_landscape}(a). Namely, the participating supporting photons can compensate for a low energy contribution $eV$ of the bias voltage by $\ell_{\rm s}\hbar\omega_{\rm s}$, thus shifting the onset of the exponential increase of $F$ to a lower bias voltage. However, the corresponding relaxation rates are suppressed by a typically small scaling factor $\rho_{\rm s}^{\left|\ell_{\rm s}\right|}$ arising from the transition matrix elements $\left|M_{kl}^{(\textrm{s})}\right|^2\propto\rho_{\rm s}^{\left|\ell_{\rm s}\right|}$ (see Appendix~\ref{sec:rfqcr}). On the other hand, the matrix elements depend on the occupation as $\left|M_{kl}^{(\textrm{s})}\right|^2\propto k^{\left|\ell_\textrm{s}\right|}$ for $k\ll 1/\rho_\textrm{s}$, which allows to sequentially activate the multiphoton processes by increasing the occupation of the supporting mode until the rates saturate at $\rho_\textrm{s}k\approx1$, beyond which the matrix elements diminish as $k^{-1/2}$. Thus, the dissipation of the primary mode can be controlled by the drive strength of the supporting mode.

\subsection{Reflection measurement}

To experimentally demonstrate the above-discussed phenomena, we probe the fundamental mode of the resonator in a standard microwave reflection measurement as illustrated in Fig.~\ref{fig:setup}(b). Ideally, the voltage reflection coefficient of a weak probe signal at the angular frequency $\omega_{\rm p,probe}$ is given by (see Appendix~\ref{sec:analysis})
\begin{equation}
	\label{eq:refl}
	\Gamma = \frac{\gamma_{\rm tr,p} - \gamma_{\rm T,p} - \gamma_{\rm 0,p} + 2{\rm i}(\omega_{\rm p,probe}-\omega_{\rm p})}{\gamma_{\rm tr,p} + \gamma_{\rm T,p} + \gamma_{\rm 0,p} - 2{\rm i}(\omega_{\rm p,probe}-\omega_{\rm p})}, 
\end{equation}
where $\gamma_{\rm tr,p}$ is the coupling strength between the resonator and the transmission line, and $\gamma_{\rm 0,p}$ is the damping rate owing to any excess sources, yielding the internal quality factor of the primary mode as $1/\gamma_{\rm 0,p}$.

Figure~\ref{fig:F_landscape}(b) shows results of reflection measurements at different probe frequencies, bias voltages, and supporting-mode drive strengths. For an increasing bias voltage or drive power, we observe broadening of the resonance dip in frequency, indicating the activation of the QCR. Here, the broadening of the full width of the dip at half maximum corresponds to stronger total coupling strength $\gamma_{\rm tot,p}=\gamma_{\rm tr,p}+\gamma_{\rm T,p}+\gamma_{\rm 0,p}$ of the system to its environments. Remarkably, the resonance frequency of the primary mode, located near the minimum reflection amplitude, exhibits clearly non-monotonic behavior which is repeated in the vicinity of each bias voltage corresponding to an onset of a different multiphoton tunnelling process. Interestingly, this shift seems to be increased at high rf powers where it may persists even in the absence of voltage bias, but Fig.~\ref{fig:F_landscape}(b) does not provide clear enough evidence for these conclusions. 

Thus for a detailed analysis, we extract $\omega_{\rm p}$, $\gamma_{\rm T,p}$, $\gamma_{\rm tr,p}$, and $\gamma_{\rm 0,p}$ by fitting the reflection coefficient to data corresponding to Fig.~\ref{fig:F_landscape}(b) (see Appendix~\ref{sec:analysis} and Fig.~\ref{fig:magnitude_fit} therein). For the corresponding theoretical model, including equation~\eqref{eq:gamma_main}, we map the rf drive power to the average supporting-mode occupation $\bar{n}_{\rm s}$ as described in Appendix~\ref{sec:pow2occ} and Fig.~\ref{fig:pow2occ} therein. In addition to the transmission line and excess losses, we take into account for both modes the loss of rf photons owing to the different types of photon-assisted electron-tunnelling events.

\subsection{On-demand dissipation}

\begin{figure*}
	\centering
	\includegraphics[width=2\columnwidth]{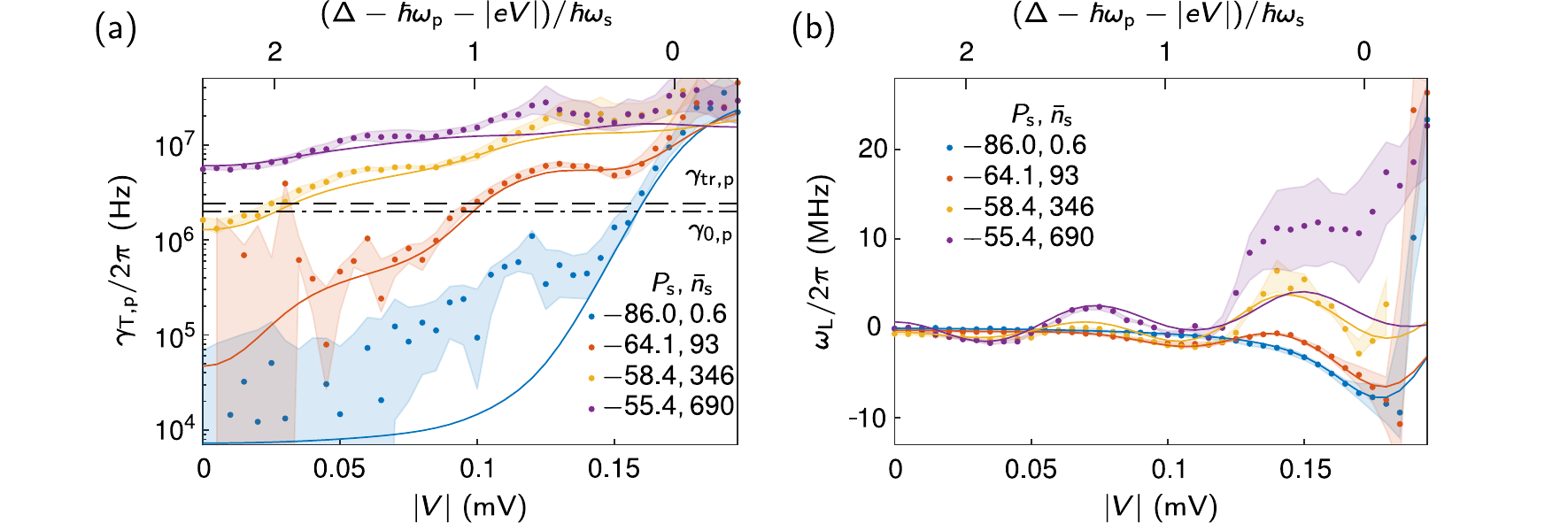}
	\caption{\label{fig:results}Coupling strength and the effective Lamb shift. Measured (dots) and modelled (solid lines) coupling strength $\gamma_{\rm T,p}$ (a) and effective Lamb shift $\omega_{\rm L}$ (b) as functions of the bias voltage at the indicated supporting-mode drive powers $P_{\rm s}$ in dBm. The estimated supporting-mode steady-state average photon number $\bar{n}_{\rm s}(V=0)$ is also shown. Top axis shows the energy needed from the supporting-mode photons to activate the QCR for the primary mode. In (a), the external coupling strength $\gamma_{\rm tr,p}$ is indicated with the dashed line and the excess coupling strength $\gamma_\textrm{0,p}$ with the dash-dotted line. The shading denotes the $1\sigma$ confidence intervals of the extracted parameters as discussed in Appendix~\ref{sec:analysis}.}
\end{figure*}

Figure~\ref{fig:results}(a) shows the measured environmental coupling strength $\gamma_{\rm T,p}$ as a function of the bias voltage at different drive powers for the supporting mode. As discussed above, with the addition of the rf excitation, the photon-assisted electron-tunnelling events can absorb photons from the supporting mode in addition to the primary-mode photons. As a result of these additional energy quanta, the coupling strength $\gamma_{\rm T,p}$ exhibits in Fig.~\ref{fig:results}(a) an exponential rise at a reduced bias voltage. We observe that with a sufficiently strong rf drive, the bias voltage is not even necessary for turning on the QCR-induced dissipation, i.e., the device is purely rf-controllable. This is beneficial in practical applications since undesired low-frequency noise can be significantly reduced by omission of dc connections to the sample. Since the rf control has a similar energy-providing effect to that of the bias voltage, it does not essentially change the maximum achievable coupling strength. Consequently, the coupling strength $\gamma_{\rm T,p}$ is tunable by three orders of magnitude, roughly from $2\pi\times10$~\si{\kilo\hertz} to $2\pi\times10$~\si{\mega\hertz}. 

Although for $\gamma_{\rm T,p} \ll \gamma_{\rm tr,p}, \gamma_{\rm 0,p}$ the measured reflection coefficient is only weakly dependent on the quantity of interest, the results are in substantial agreement with our theoretical model of the rf QCR as demonstrated in Fig.~\ref{fig:results}(a). However, note that the supporting-mode attenuation has been used as a fitting parameter as well as a single electron temperature $T_{\rm N}$ that is independent of the bias voltage or drive power (see Appendices~\ref{sec:measurement} and~\ref{sec:analysis}). We expect a more comprehensive study of the electron temperature to explain in the future most of the differences between the model and the measurements, such as the interesting behavior observed at high powers where the oscillations of the experimentally observed damping are are much more pronounced than those of the model. This difference cannot be explained by typical experimental noise which tends to smoothen such oscillations.

\subsection{Photon-number-dependent effective Lamb shift}

Figure~\ref{fig:results}(b) shows the observed shift in the primary-mode frequency $\omega_{\rm L} = \omega_{\rm p} - \omega_{\rm p}^0$ as a function of the bias voltage and of the rf drive strength, where $\omega_{\rm p}^0$ is the frequency for the QCR turned off. We observe that the shift strongly depends on the rf drive strength, and thus on the supporting-mode photon number. Since the linear resonator mode exhibits no ac Stark shift by the environment, we conclude that the observed shift is an effective Lamb shift that depends on the photon number in the engineered hybrid environment through the photon number dependence of the coupling strength of Eq.~\eqref{eq:gamma_main}. The fermionic part of the hybrid environment, i.e., electron tunnelling, is required to mediate the interaction between the bosonic system and the bosonic part of the environment.

The effective Lamb shift in Fig.~\ref{fig:results}(b) exhibits an oscillation for each multi-photon-assisted electron tunnelling process, changing between \num{-10}~\si{\mega\hertz} to \num{22}~\si{\mega\hertz} with the bias voltage. Importantly, under pure rf control the observed effective Lamb shift lies between \num{-0.6}~\si{\mega\hertz} to \num{0}~\si{\mega\hertz}. 
Thus in the case of a purely rf-controlled QCR, fluctuations in the control parameters of the QCR may lead to less spurious dephasing of the coupled quantum system than for a dc-controlled QCR. Furthermore, the rf-controlled effective Lamb shift appears as a non-linear interaction between the two resonator modes which is a key constituent for using otherwise linear resonators as qubits.

Next, we validate our claim that the observed frequency shift is in fact the effective Lamb shift by comparing our experimental observations with a first-principles theoretical model which we derive in Appendix~\ref{sec:lamb}. For the dynamic effective Lamb shift, which is the full environment-induced frequency shift with the static, zero-frequency component subtracted, we obtain  
\begin{equation}
	\label{eq:lamb_dyn}
	\hspace{-.2cm} 
	\omega_{\rm L} = -{\rm PV}\int\limits_0^\infty\frac{{\rm d}\omega}{2\pi}\left[\frac{\gamma_{\rm T,p}(\omega)}{\omega-\omega_{\rm p}^0} + \frac{\gamma_{\rm T,p}(\omega)}{\omega+\omega_{\rm p}^0} - 2\frac{\gamma_{\rm T,p}(\omega)}{\omega}\right], 
\end{equation}
where PV is the Cauchy principal value integration and the coupling strength $\gamma_{\rm T,p}(\omega)$ is given by equation~\eqref{eq:gamma_main} such that $V$ and $\bar{n}_{\rm s}$ are considered independent of $\omega=\omega_\textrm{p}$. In addition, the fundamental frequency of a harmonic oscillator undergoes a classical damping shift of $\sim\gamma_{\rm T,p}^2/(8\omega_{\rm p})$, although in our system this is only of the order of \num{10}~\si{\kilo\hertz}. Furthermore, we observe a static shift of $-\mu\gamma_{\rm T,p}/\pi$ with a proportionality constant $\mu=0.64$ at the two lowest supporting-mode drive strengths. We attribute this shift to an effective elongation of the resonator mode~\cite{Silveri19}, arising from the exponentially increased tunnelling rate which leads to a decreased effective junction impedance, and hence increased current flow through the junction. We do not claim the existence of a static shift at the higher drive strengths owing to the experimental uncertainties. The model describes well the experimental observations, as shown in Fig.~\ref{fig:results}(b), except for the large measured effective Lamb shift at the highest rf drive strengths and bias voltages. This discrepancy is consistent with the differences between the measured and the modeled coupling strengths visible in Fig.~\ref{fig:results}(a). These differences are partly explained by multiphoton events involving only the supporting mode, a phenomenon which is possible to include in our model, however not considered here.

In the reminder of this section, let us discuss possible alternative explanations for the observed frequency shift. Fortunately, we find that such alternatives are not in line with the experimental data, thus supporting the validity of the above-given interpretation. 

The frequency shift arising from the transformation of linearly coupled harmonic oscillators into the corresponding normal modes is equal for the quantum and classical models. Occasionally, this type of a contribution, referred to as normal-mode splitting, is not considered to be a part of the Lamb shift~\cite{Gely18}. In our discussion above, we did not explicitly differentiate between these. However, since we measure the effective Lamb shift with respect to the zero-bias zero-drive-power case, our result is free from any normal-mode splitting that may occur owing to the presence of the modes of the resonator. Furthermore, the measured effective Lamb shift is mediated by electron tunneling with no direct coupling of the primary mode to additional physical bosonic modes. Thus, we do not consider the classical normal-mode splitting to significantly contribute to our observations.

\begin{figure}
	\centering
    \includegraphics[width=\columnwidth]{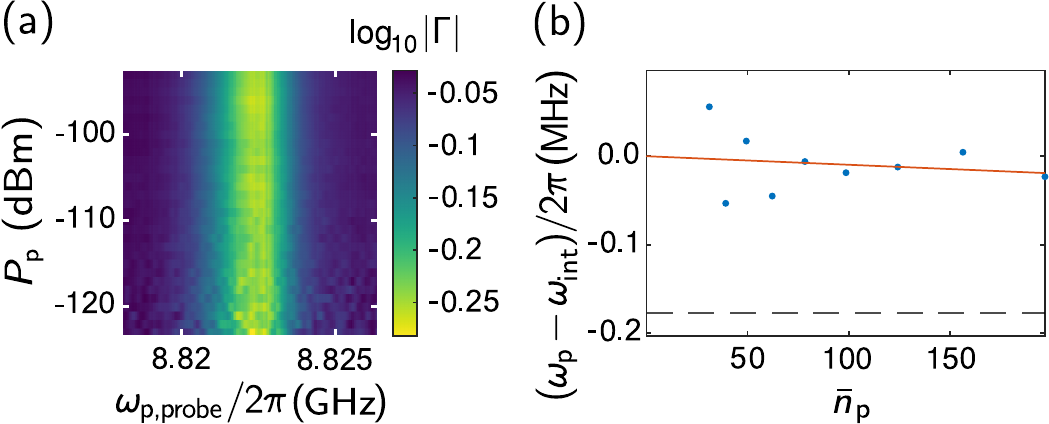}
    \caption{\label{fig:kerr} Self-Kerr. (a) Experimentally measured reflection magnitude (not normalized) as a function of the probe frequency and the probe power. The data are taken at $P_{\rm s}=-92.4$~dBm and $V=0.095$~mV. At the resonance, $P_{\rm p} = -116$~dBm corresponds to an average primary-mode photon occupation $\bar{n}_{\rm p}$ of the order of $1$. (b) Resonance frequency extracted from the reflection coefficient (dots) as a function of the occupation $\bar{n}_{\rm p}$. The dashed line indicates the theoretically predicted Lamb-shifted resonance frequency. The slope of the linear fit (red) with an intercept $\omega_{\rm int}/2\pi = 8.82230$~GHz ($1\sigma$ uncertainty $0.025$~MHz) yields estimated self-Kerr coefficient $-0.1$~kHz ($1\sigma$ uncertainty $0.25$~kHz). The intercept of the linear fit bears an additional systematic uncertainty of approximately $0.2$~MHz which however does not affect the slope.}
\end{figure}

\begin{figure}
	\centering
    \includegraphics[width=\columnwidth]{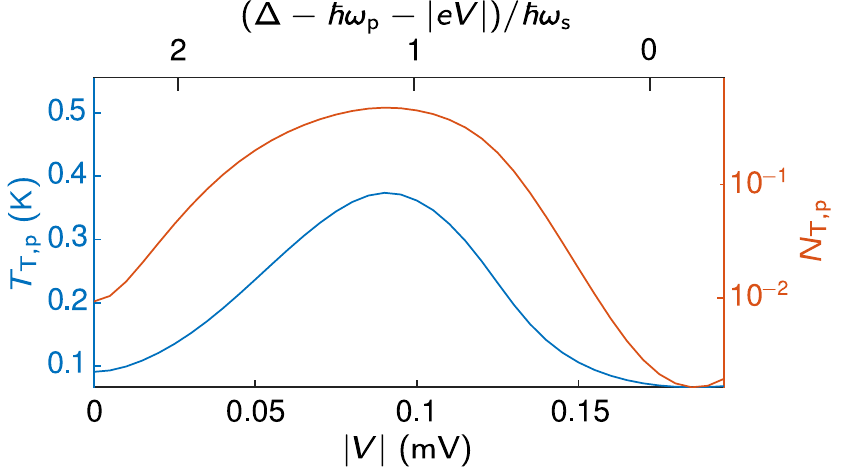}
    \caption{\label{fig:temperature} Characteristics of the effective electromagnetic environment formed by the QCR. Temperature (blue line, left axis) and the corresponding thermal occupation (orange line, right axis) of the QCR environment as functions of the bias voltage at the supporting-mode drive power $P_{\rm s}=-118.4$~dBm. Top axis shows the energy needed from the supporting-mode photons to activate the QCR for the primary mode.}
\end{figure}

To study possible contributions of the typical ac Stark shift on our results, we have made an attempt to measure in a standard microwave reflection experiment the strength of the self-Kerr effect in the primary mode as a function of probe frequency and power as shown in Fig.~\ref{fig:kerr}. We find no self-Kerr effect within the experimental uncertainty of well below 0.1~MHz for hundred photons in the primary mode. In Fig.~\ref{fig:results}(b) in contrast, hundred photons in the supporting mode lead to a megahertz-level shift at the considered bias point. Since both modes are coupled in a similar fashion to the NIS junction, the strengths of any self-Kerr and cross-Kerr effects promoted by the junction are expected to be roughly equal. Consequently, our experimental data does not support the cross-Kerr effect as an origin of the observed frequency shifts. In addition, any ac Stark shift arising from the cross-Kerr terms is monotonic as a function of photon number, $\propto\bar{n}_\textrm{s}$, whereas the observed shift in Fig.~\ref{fig:results}(b) between $0.05$--$0.1$~mV is not. Thus, we conclude that the ac Stark shift of the primary mode, arising from the photon occupation of the secondary mode, has likely a negligible contribution to the measured frequency shifts. Furthermore, in this bias range the effective temperature of the QCR environment is the largest, see Fig.~\ref{fig:temperature},
whereas the measured frequency shifts are on the smallest end. Thus, the observed shift is not measurably dependent on the effective population of the QCR environment either, and hence we may also rule out the possibility that the observed effective Lamb shift could be modelled as an ac Stark shift arising from the effective electromagnetic environment induced by the QCR.

Furthermore, since the resistance of the NIS junction is non-linear, it may give rise to a classical-like frequency shift which changes as a function of the bias voltage and drive power of the supporting mode. This consideration includes the case of frequency mixing between the two modes, the frequencies of which differ roughly by a factor of two. However, since the differential conductance of an NIS junction has less than three minima and maxima for positive bias voltages and we observe three local maxima and minima of the frequency shift in Fig.~\ref{fig:results}(b) such an explanation for the phenomenon responsible of our results is ruled out.

Interestingly, our findings are qualitatively supported by a rather general theoretical approach to the Lamb shift from hybrid environments with initial results derived in Appendix~\ref{sec:lamb_general}. We aim to develop this theory further in the future.

\section{Discussion}

In conclusion, we demonstrated and quantitatively modelled dc and rf control of the coupling strength between a resonator mode and its electromagnetic environment formed by photon-assisted electron tunnelling in NIS junctions. We found the dissipation to be tunable by three orders of magnitude using only rf driving. Peculiarly, we observed a photon-number-dependent effective Lamb shift of the linear resonator mode which oscillates as a function of the bias voltage owing to the staggered onset of the multi-photon absorption processes. Each process strengthens the coupling of the bosonic mode to the environment, resulting in the oscillation in the coupling strength and effective Lamb shift. Importantly, using tunable rf power at vanishing bias voltage the dissipation strength can be controlled with almost vanishing effective Lamb shift. Our device concept provides opportunities for rf-controlled rapid on-demand initialization of quantum circuits, such as qubits~\cite{Sevriuk2019,Hsu20}. We envision the rf QCR to become an important component in the emerging field of quantum computing and technology.

\begin{acknowledgments}
This research was financially supported by the European Research Council under grant no.~681311 (QUESS) and Marie Sk\l{}odowska-Curie grant no.~795159; by the Academy of Finland under its Centres of Excellence Program grant nos.~312300, 336810 and 312059 and grant nos.~265675, 305237, 305306, 308161, 314302, 316551, 316619, and 318937; and by the Alfred Kordelin Foundation, the Emil Aaltonen Foundation, the Vilho, Yrj\"o and Kalle V\"ais\"al\"a Foundation, the Jane and Aatos Erkko Foundation, and the Technology Industries of Finland Centennial Foundation. We thank the provision of facilities and technical support by Aalto University at OtaNano~\---~Micronova Nanofabrication Centre.
\end{acknowledgments}

\appendix

\section{Sample fabrication} \label{sec:fabrication}

\begin{figure*}
	\centering
    \includegraphics[width=2\columnwidth]{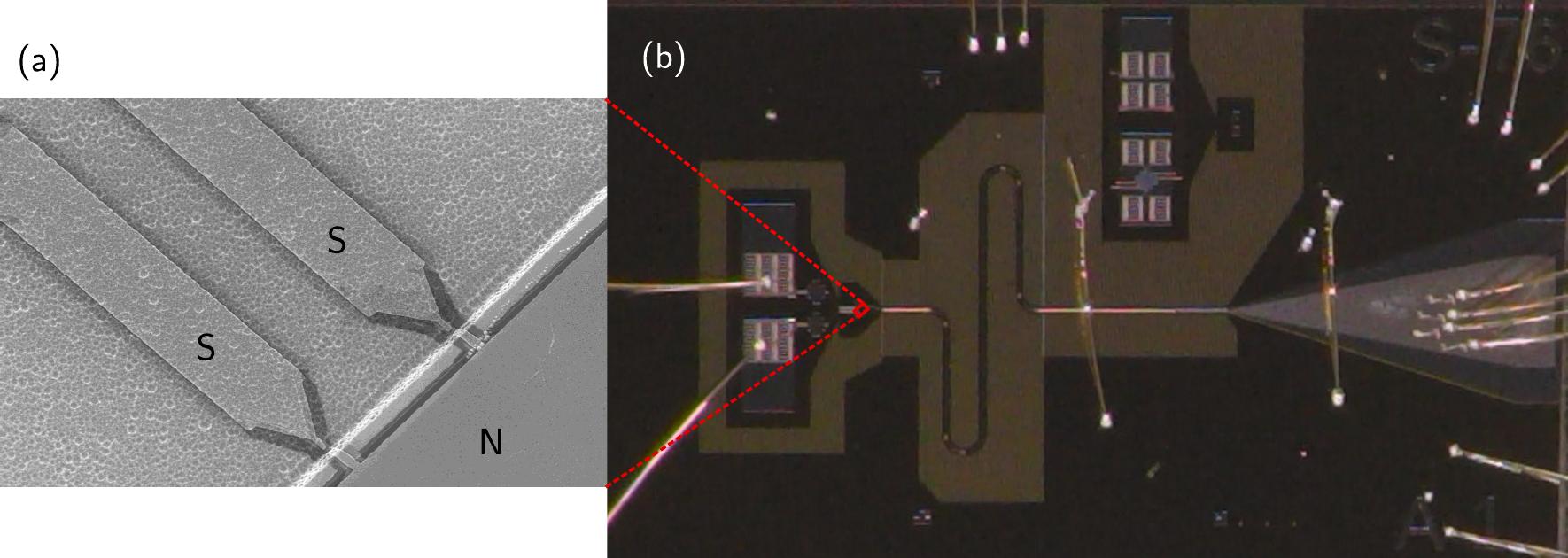}
    \caption{\label{fig:sampleimage}Sample images. (a) Scanning electron micrograph of the superconductor--insulator--normal-metal tunnel junctions of a quantum-circuit refrigerator. The scale bar denotes 5~$\mu$m. (b) Optical image of a device similar to the the one used in this study. The tunnel junctions (red box) are connected to the coplanar-waveguide resonator (winding line), which is excited through the transmission line (broadening line). The dc junction bias is applied through \textit{RC} filters between the bonding pads and the junctions. The test structures shown at the top middle of the image are not relevant for the operation of the device. The width of the field of view is 7.5~mm.}
\end{figure*}

We fabricate the sample, see Fig.~\ref{fig:sampleimage}, on a high-purity \num{500}-\si{\micro\metre}-thick silicon wafer passivated with a thermally-grown silicon oxide layer of \num{300}-\si{\nano\metre} thickness. The coplanar waveguide resonator is patterned with photolithography and reactive ion etching in a \num{200}-\si{\nano\metre}-thick sputtered niobium layer. A \num{50}-\si{\nano\metre}-thick dielectric layer of Al\textsubscript{2}O\textsubscript{3} is  atomic-layer deposited at \SI{200}{\celsius} on the entire wafer to operate as the insulator in the parallel plate capacitors. The NIS junctions are defined with electron-beam lithography and two-angle evaporation as follows: First, a \num{20}-\si{\nano\metre}-thick superconducting Al layer is evaporated, followed by in-situ oxidation to form the tunnel barriers. Second, a \num{20}-\si{\nano\metre}-thick normal-metal Cu layer is evaporated. Further fabrication details are discussed in ref.~\cite{Masuda2018}.

\section{Measurements} \label{sec:measurement}

We cool the sample down to a base temperature of \SI{10}{\milli\kelvin} in a commercial dilution refrigerator. The sample is attached to a gold-plated copper sample holder with a printed circuit board, to which the sample is bonded with aluminum wires.

We measure the reflection coefficient of the sample with a vector network analyzer (VNA) at frequencies close to that of the primary mode of the CPW resonator. We drive the supporting mode with a microwave signal generator through the same port as we use for the VNA, as illustrated in Fig.~\ref{fig:setup}b. The VNA tone is attenuated to the single-photon regime before it is introduced to the sample. The rf drive power for the supporting mode is controlled by three orders of magnitude. The rf tone attenuation of the primary mode, $-103$~dB, is obtained by separately measuring the room temperature components and using the manufacturer specifications for the cryogenic components. For the supporting mode, the frequency range is outside that specified for some cryogenic components, and consequently we use $-105$~dB obtained as a fitting parameter.  
The superconductor--insulator--normal-metal--insulator--superconductor (SINIS) junction is biased with a ground-isolated room temperature voltage source.

\section{Data analysis} \label{sec:analysis}

\begin{figure}
	\centering
	\includegraphics[width=\columnwidth]{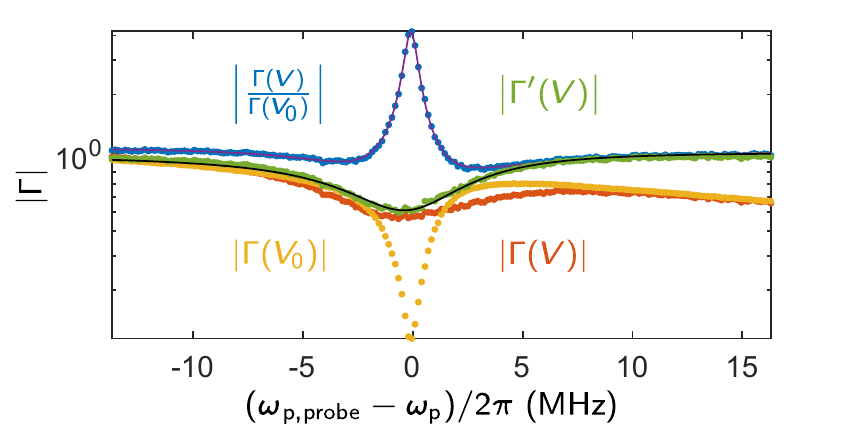}
	\caption{\label{fig:magnitude_fit}Reflection data analysis. Reflection magnitude $|\Gamma|$ as a function of the probe frequency $\omega_{\rm p,probe}$ at two different bias voltages and supporting mode powers: $V = 0.08~\si{\milli\volt}, P_{\rm s} = -90.8~\mathrm{dBm}$ (red) and $V_0 = 0~\si{\milli\volt}, P_{0} = -118.4~\mathrm{dBm}$ (yellow). Fits to several fractions of two reflection magnitudes such as $r=|\Gamma(V,P_{\rm s})/\Gamma(V_0,P_{0})|$ (blue) are used to extract an averaged off-state reflection magnitude $\Gamma^\prime(V_0,P_{0})$. Example of background subtracted reflection magnitude (green) is obtained as $|\Gamma^\prime(V,P_{\rm s})|=|r\Gamma^\prime(V_0,P_{0})|$.}
\end{figure}

We subtract the background in the measured VNA frequency traces as described in ref.~\cite{Hyyppa19} and illustrated in Fig.~\ref{fig:magnitude_fit}. We discuss the method here because of the additional tuning parameter, the supporting mode excitation. Namely, we divide the raw measured reflection coefficient $\Gamma(V,P_{\rm s})$ by the off-state reflection trace $\Gamma(0,P_0)$ and fit a double Lorentzian of the form $r=\frac{\Gamma(V,P_{\rm s})}{\Gamma(0,P_0)}$, where each reflection coefficient $\Gamma$ is given by 
\begin{equation}
	\label{eq:refl_fano}
	\Gamma = \frac{2\gamma_{\rm tr,p} - r_0[\gamma_{\rm tr,p} + \gamma_{\rm T,p} + \gamma_{\rm 0,p} - 2{\rm i}(\omega_{\rm p,probe}-\omega_{\rm p})]}{\gamma_{\rm tr,p} + \gamma_{\rm T,p} + \gamma_{\rm 0,p} - 2{\rm i}(\omega_{\rm p,probe}-\omega_{\rm p})}, 
\end{equation}
where $r_0$ is a complex-valued Fano factor~\cite{Fano61} such that $|r_0|=1$. This procedure is carried out for each voltage $V$ and rf drive strength $P_{\rm s}$, yielding a background-subtracted off-state reflection trace $\Gamma^\prime(0,P_0)$ which is determined by equation~\eqref{eq:refl_fano} with the off-state fit parameters averaged over those obtained from the fits for different $V$ and $P_{\rm s}$. We retrieve the background-corrected traces as $\Gamma^\prime(V,P_{\rm s})=r\Gamma^\prime(0,P_0)$ and fit equation~\eqref{eq:refl_fano} to the obtain the results for the primary-mode frequency $\omega_{\rm p}$, the coupling strength $\gamma_{\rm T,p}$, the external coupling strength $\gamma_{\rm tr,p}$, and the excess coupling strength $\gamma_\textrm{0,p}$ used in this manuscript.

Ideally, in the absence of the Fano correction ($r_0=1$), the reflection coefficient vanishes at the critical points where the impedance of the transmission line and of the resonator are matched at $\omega_{\rm p,probe} = \omega_{\rm p}$ and $\gamma_{\rm tr,p} = \gamma_{\rm T,p} + \gamma_{0,\textrm{p}}$. For these points, the full width of the dip, $2\gamma_{\rm tr,p}$, yields accurately the coupling strength to the transmission line. However, a non-vanishing Fano correction may shift the dip and make it shallower, although in our measurements the correction is small. Note that the critical points can also be identified from the phase of reflection coefficient as it exhibits a full $2\pi$ phase winding about the critical points. The damping rate owing to excess losses $\gamma_{0,\textrm{p}}$ is extracted at zero voltage and minimum supporting-mode power where the QCR is effectively decoupled from the rest of the circuit, i.e., $\gamma_{\rm T,p}$ is negligible. Note that the measurements for $\gamma_{\rm T,p}\ll\gamma_{\rm tr, p},\gamma_{\rm 0, p}$ are challenging since the measured quantity is essentially a result of a subtraction of several other large quantities, thus amplifying the relative uncertainty of the result. Consequently, our results may be unreliable at such a regime, which is also reflected as relatively large experimental uncertainty bounds obtained as described below.

We obtain the $1\sigma$ confidence intervals of each extracted parameter of equation~\eqref{eq:refl_fano} individually. In this method~\cite{Silveri19}, we vary the parameter while the other parameters correspond to the optimal least-squares fit and calculate the resulting resonance point given by equation~\eqref{eq:refl_fano}. The $1\sigma$ confidence interval of the parameter is determined by the condition that the distance of the calculated resonance point from that of the least-squares fit in the complex plane is smaller or equal to the root-mean-square fit error of the reflection coefficient.

The quasiparticle temperature of the superconducting leads and the electron temperature of the normal-metal island are higher than the base temperature of the refrigerator, for example, owing to heat leakage through the radiation shields and to heating from the input signals. Note that such low heating power does not affect the temperature of the macroscopic dilution refrigerator as much as the temperature of the small normal-metal island which is only weakly thermally coupled to the dilution refrigerator. The temperature affects the Fermi distributions and determines the slope of the exponential growth in the coupling strength as a function of the bias voltage for the onset of each multiphoton process, but not the coupling strength, to which each photon-assisted electron tunnelling process saturates to. Although in general the temperature is also affected by the amount and type of quasiparticle tunneling processes, we observed that our data is fairly well described by a single bias and power-independent quasiparticle temperature.
This is likely because the temperature has a noticeable broadening effect only on steep slopes, which occur only at low input powers and at short voltage ranges. Thus, although the temperature is in principle input power and voltage dependent in the measurement range~\cite{Masuda2018}, we use for simplicity a single value of the quasiparticle temperature $T_{\rm N}$ in Table~1 for all measurement points, obtained by the best fit of the theoretical model to the curve in Fig.~\ref{fig:results} with the lowest supporting-mode power. Fitting the temperature individually for each measurement point would result in an artificially good match between the theory and data. Expanding on this procedure, the device may be used as a thermometer. In this case, the device parameters may be fitted at the slope of $\gamma_{\rm T,p}$ as a function of voltage at the lowest supporting-mode drive strength. Then, the rest of the data may be fitted with $T_{\rm N}$ as the sole free parameter.
Further details of parameter extraction can be found in ref.~\cite{Silveri19}.

\section{Photon-assisted electron tunnelling} \label{sec:rfqcr}

Following the approach of ref.~\cite{Silveri17} for a single mode, we derive the coupling strength between the environment and the resonator based on the lumped-element circuit model given in Fig.~\ref{fig:circuit} for the two modes.

We begin by assuming that the two modes are far off resonance with each other, and thus the effect of their capacitive interaction is negligible to the lowest order. Our quantum-mechanical circuit is thus described by the Hamiltonian~\cite{Wallraff04,Blais04} 
\begin{align}
    \hat{H} =& \frac{(\hat{Q}_{\rm p}+\alpha_{\rm p}\hat{Q}_{\rm N})^2}{2C^\prime_{\rm p}} + \frac{(\hat{Q}_{\rm s}+\alpha_{\rm s}\hat{Q}_{\rm N})^2}{2C^\prime_{\rm s}} \nonumber \\*
    &+ \frac{\hat{Q}_{\rm N}^2}{2C^\prime_{\rm N}} + \frac{\hat{\Phi}^2_{\rm p}}{2L_{\rm p}} + \frac{\hat{\Phi}^2_{\rm s}}{2L_{\rm s}}, 
\end{align}{}
where $\hat{Q}_{\rm p}$, $\hat{Q}_{\rm s}$, and $\hat{Q}_{\rm N}$ are the conjugate charge operators of the primary and secondary resonator modes and the normal-metal island, respectively, $\hat{\Phi}_{\rm p}$ and $\hat{\Phi}_{\rm s}$ are the respective node flux operators, and the capacitance ratios and the renormalized capacitances are given by
\begin{equation}
    \label{eq:capacitance_ratio}
    \alpha_{\rm p} = \frac{C_{\rm c,p}(C_{\rm c,s}+C_{\rm s})}{C_{\rm c,s}(C_{\rm c,p}+C_\Sigma)+C_{\rm s}(C_{\rm c,p}+C_{\rm c,s}+C_\Sigma)},
\end{equation}{}
\begin{equation}
    C^\prime_{\rm p} = C_{\rm p}+\alpha_{\rm p}\left(C_\Sigma+\frac{C_{\rm s}C_{\rm c,s}}{C_{\rm c,s}+C_{\rm s}}\right),
\end{equation}{}
\begin{widetext}
\begin{equation}
    \label{eq:CN}
	C_{\rm{N}}' = \frac{C_{\rm c,s}(C_{\rm p}+C_{\rm c,p})C_{\rm s}+(C_{\rm p}+C_{\rm c,p})(C_{\rm s}+C_{\rm c,s})C_\Sigma+C_{\rm c,p}(C_{\rm s}+C_{\rm c,s})C_{\rm p}}{-C_{\rm c,s}(C_{\rm p}+C_{\rm c,p})\alpha_{\rm s}+(C_{\rm p}+C_{\rm c,p})(C_{\rm s}+C_{\rm c,s})-C_{\rm c,p}(C_{\rm s}+C_{\rm c,s})\alpha_{\rm p}},
\end{equation}
\end{widetext}
where $C_\Sigma = C_{\rm m} + C_{\rm j}$ is the total capacitance of the normal-metal island and $\alpha_{\rm s}$ and $C^\prime_{\rm s}$ are obtained by swapping the subscripts $\rm{s}$ and $\rm{p}$ in the above equations.

The tunnelling events are considered as weak perturbations inducing transitions between the eigenstates $\ket{m_{\rm p/s}}$ and $\ket{m^\prime_{\rm p/s}}$ of each resonator mode and are characterized by the transition matrix elements 
\begin{align}
	\label{eq:M_app}
	&\left|M_{mm'}^{({\rm p/s})}\right|^2 \nonumber \\*
	&= \textrm{e}^{-\rho_{\rm p/s}}\rho_{\rm p/s}^{\left|\ell\right|}\left(\frac{m'!}{m!}\right)^{\mathrm{sgn}(\ell)}\left|L_{\min(m,m')}^{\left|\ell\right|}(\rho_{\rm p/s})\right|^2, 
\end{align}
where $\rho_{\rm p/s}=\frac{\pi\alpha_{\rm p/s}^2}{\omega_{\rm p/s} C_{\rm p/s}R_{\rm K}}$, $\ell=m-m'$, $R_{\rm K}=\frac{h}{e^2}$ is the von Klitzing constant, and $L_{\min(m,m')}^{\left|\ell\right|}(\rho_{\rm p/s})$ is the generalized Laguerre polynomial~\cite{Abramowitz72}.

We assume the two NIS junctions to be identical, the electrodes to be at equal temperature $T_\textrm{N}$, and the charging energy $E_{\rm N} = e^2/(2C_{\rm N}^\prime)$ of the normal-metal island to be much smaller than other relevant energy scales, i.e., $E_{\rm N}\ll \hbar\omega_{\rm p/s},\Delta,k_{\rm B}T_{\rm N}$, where $\Delta$ is the Bardeen--Cooper--Schrieffer gap parameter. Fermi's golden rule describes the transition rates between the resonator number states as 
\begin{align}
	\Gamma_{(m_{\rm p},m_{\rm s}),(m_{\rm p}',m_{\rm s}')}&(V) \nonumber \\*
	=& \left|M_{m_{\rm p}m_{\rm p}'}^{({\rm p})}\right|^2\left|M_{m_{\rm s}m_{\rm s}'}^{({\rm s})}\right|^2\frac{2R_{\rm{K}}}{R_{\rm{T}}}\sum_{\tau=\pm1} \nonumber \\*
	&F(\tau eV+\ell_{\rm p}\hbar\omega_{\rm p} + \ell_{\rm s}\hbar\omega_{\rm s}-E_{\rm{N}}), 
\end{align}
where $R_{\rm T}$ is the tunnelling resistance. Here, we denote by $F(E)$ the normalized forward tunnelling rate at the bias energy $E$ 
\begin{equation}
    \label{eq:F}
	F(E) = \frac{1}{h}\int \mathrm{d}\varepsilon\mathinner n_{\rm{S}}(\varepsilon)[1-f_{\rm{S}}(\varepsilon)]f_{\rm{N}}(\varepsilon-E), 
\end{equation}
where $f_{\rm S}$ and $f_{\rm N}$ are the Fermi distributions of the superconductor and of the normal metal, respectively. The function $n_{\rm S}$ is the normalized quasiparticle density of states in the superconductor~\cite{Dynes78} 
\begin{equation}
	n_{\rm{S}}(\varepsilon) = \left|\rm{Re}\left\{\frac{\varepsilon+i\gamma_{\rm{D}}\Delta}{\sqrt{(\varepsilon+i\gamma_{\rm{D}}\Delta)^2-\Delta^2}}\right\}\right|, 
\end{equation}
where $\gamma_{\rm D}$ is the Dynes parameter.

The interaction parameter of the primary resonator $\rho_{\rm p}$ is typically small, and thus at low powers, transitions between adjacent primary-mode states $\Gamma_{(m_{\rm p},m_{\rm s}),(m_{\rm p} \pm 1,m_{\rm s}')}$ are dominant, see equation~\eqref{eq:M_app}. We obtain the coupling strengths of the individual resonators by tracing out the other resonator as, for example, for the primary mode $\widetilde{\Gamma}_{mm^\prime} = \sum_{k,l}P_k\Gamma_{(m,k),(m^\prime,l)}$, where $P_k = \textrm{e}^{-\bar{n}_{\rm s}}\frac{\bar{n}_{\rm s}^k}{k!}$ is the likelihood that the supporting mode Fock state $\ket{k}$ is occupied due to a classical drive and $\bar{n}_{\rm s}$ is the mean supporting-mode photon number, and by noting the relations between the transition rates and the coupling strength, $\gamma_{\rm T,p}=\widetilde{\Gamma}_{10}-\widetilde{\Gamma}_{01}$, and the effective mode temperature, $T_{\rm T,p}=\frac{\hbar\omega_{\rm p}}{k_{\rm B}}\left[\ln{\left(\frac{\widetilde{\Gamma}_{10}}{\widetilde{\Gamma}_{01}}\right)}\right]^{-1}$. Consequently, the effective electromagnetic environment of the primary mode is characterized by its coupling strength 
\begin{widetext}
\begin{equation}
	\label{eq:gamma}
	\gamma_{\rm T,p}(V,\bar{n}_{\rm s}) = 2\pi\alpha_{\rm p}^2\frac{Z_{\rm p}}{R_{\rm{T}}}\sum_{k,l}P_k(\bar{n}_{\rm s})\left|M_{kl}^\textrm{(s)}\right|^2\sum_{\ell_{\rm p},\tau=\pm1}\ell_{\rm p} F(\tau eV+\ell_{\rm p}\hbar\omega_{\rm p}+\ell_{\rm s}\hbar\omega_{\rm s}-E_{\rm{N}}), 
\end{equation}
where $\ell_\textrm{s}=k-l$ and the effective mode temperature
\begin{equation}
    T_{\rm T,p}(V,\bar{n}_{\rm s}) = \frac{\hbar\omega_{\rm p}}{k_{\rm B}}\left[\ln{\left(\frac{\sum_{k,l}P_k\left|M_{kl}^\textrm{(s)}\right|^2\sum_{\tau=\pm1}F(\tau eV+\hbar\omega_{\rm p}+\ell_{\rm s}\hbar\omega_{\rm s}-E_{\rm N})}{\sum_{k,l}P_k\left|M_{kl}^\textrm{(s)}\right|^2\sum_{\tau=\pm1}F(\tau eV-\hbar\omega_{\rm p}+\ell_{\rm s}\hbar\omega_{\rm s}-E_{\rm N}}\right)}\right]^{-1}.
\end{equation}{}
\end{widetext}
The average thermal occupation of the environment at the temperature $T_{\rm T,p}$ is given by $N_{\rm T,p} = 1/\{\exp{[\hbar\omega_{\rm p}/(k_{\rm B}T_{\rm T,p})]}-1\}$.

In addition to the coupling strength, we have derived equation~\eqref{eq:lamb_dyn} for the effective Lamb shift using second-order perturbation theory for the eigenfrequencies of the primary mode. This derivation is presented in Appendix~\ref{sec:lamb}.

\section{Mapping of input power to occupation} \label{sec:pow2occ}

\begin{figure}
	\centering
    \includegraphics[width=\columnwidth]{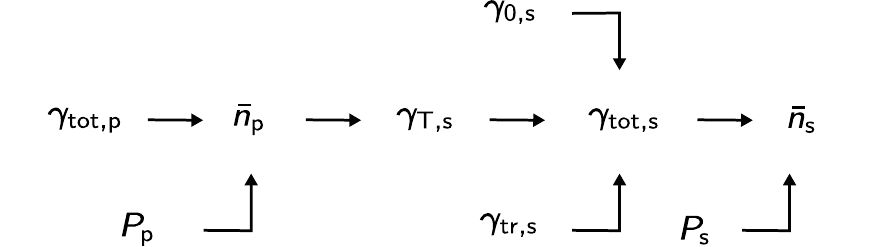}
    \caption{\label{fig:pow2occ}Extraction of the supporting-mode occupation. The primary mode is driven at strength $P_{\rm p}$, which due to the total coupling strength $\gamma_{\rm tot,p}$ results in a steady-state occupation $\bar{n}_{\rm p}$. This occupation affects the coupling strength of the supporting mode to the QCR, $\gamma_{\rm T,s}$, which together with the external $\gamma_{\rm tr,s}$ and excess coupling $\gamma_{\rm 0,s}$ of the mode is the total coupling strength $\gamma_{\rm tot,s}$. The steady-state occupation $\bar{n}_{\rm s}$ is obtained from $\gamma_{\rm tot,s}$ and the supporting mode drive strength $P_{\rm s}$ using Eq.~\eqref{eq:n_steady}.}
\end{figure}

The driven supporting mode is modelled by the rf drive Hamiltonian in the rotating frame
\begin{equation}
	\hat{H}_{\rm drive} = \hbar(\Omega_ \textrm{s}\hat{a} + \Omega_ \textrm{s}^*\hat{a}^\dagger), 
\end{equation}
where $\Omega_ \textrm{s}$ is the complex-valued rf drive amplitude as the resulting Rabi angular frequency and $\hat{a}$ and $\hat{a}^\dagger$ are the annihilation and creation operators of the supporting mode. The combined effect of the drive and dissipation results in a steady-state occupation of 
\begin{equation}
	\label{eq:n_steady}
    \bar{n}_{\rm s}= \frac{|\Omega_{\textrm{s}}|^2}{\gamma_{\rm tot,s}^2+\Delta_{\rm s}^2} = \frac{2\omega_{\rm s}^2}{\pi\hbar(\gamma_{\rm tot,s}^2+\Delta_{\rm s}^2)}P_{\rm s}C_{\rm g,s}^2Z_{\rm tr,s}Z_{\rm s},
\end{equation}
where $\Delta_{\rm s}$ is the detuning of the rf drive, $Z_{\rm s}$ is the characteristic impedance of the supporting mode and $P_{\rm s}$ is the rf drive strength at the input capacitor of size $C_\textrm{g,s}$. Here, $\gamma_{\rm tot,s} = \gamma_{\rm T,s} + \gamma_{\rm 0,s} + \gamma_{\rm tr,s}$ is the total coupling strength of the supporting mode and consists of the coupling of the mode to the QCR $\gamma_{\rm T,s}$, to external sources $\gamma_{\rm tr,s}$, and to excess sources $\gamma_{\rm 0,s}$. We assume the damping rate due to excess sources $\gamma_{\rm 0,s}$ to have a minor effect here since the external coupling is much higher than it. The coupling strength of the supporting mode to the transmission line is calculated as~\cite{Silveri17} 
\begin{equation}
	\gamma_{\rm tr,s} = \frac{Z_{\rm s}}{Z_{\rm tr,s}}\frac{\omega_{\rm s}^3}{\omega_{\rm s}^2+\omega_{\rm RC,s}^2}, 
\end{equation}
where $\omega_{\rm RC,s} = 1/(Z_{\rm tr,s}C_{\rm g,s})$ is the corresponding angular frequency. The resonator frequency $\omega_{\rm s}/(2\pi)$ originates from 
\begin{equation}
	\omega_{\rm s}^\prime \approx \omega_{\rm s}-\frac{Z_{\rm s}}{2Z_{\rm tr,s}}\frac{\omega_{\rm s}^2\omega_{\rm RC,s}}{\omega_{\rm s}^2+\omega_{\rm RC,s}^2}, 
\end{equation}
where $\omega_{\rm s}^\prime/(2\pi)$ is the renormalized resonator frequency which is experimentally measured and employed as the rf drive frequency. The coupling strength between the supporting mode and the QCR, $\gamma_{\rm T,s}$, is obtained from equation~\eqref{eq:gamma} by exchanging the roles of the primary and supporting modes. Furthermore, the steady-state occupation of the primary mode $\bar{n}_{\rm p}$ is calculated with equation~\eqref{eq:n_steady} by substituting the measured values of $\gamma_{\rm tot,p}$, the primary-mode drive strength $P_{\rm p}$, and the detuning of the primary-mode drive $\Delta_{\rm p}$. The procedure is illustrated in Fig.~\ref{fig:pow2occ}.

In the future, we intend to improve the model by iteration of the above procedure and alternatively, by replacing it with solving 
\begin{equation}
\begin{cases}
    0 = -i\Delta_{\rm s}(\bar{n}_{\rm p})\bar{\alpha}_{\rm s} - \frac{\gamma_{\rm tot,s}(\bar{n}_{\rm p})}{2}\bar{\alpha}_{\rm s} + i\Omega_{\rm s} \\
    0 = -i\Delta_{\rm p}(\bar{n}_{\rm s})\bar{\alpha}_{\rm p} - \frac{\gamma_{\rm tot,p}(\bar{n}_{\rm s})}{2}\bar{\alpha}_{\rm p} + i\Omega_{\rm p}, 
\end{cases}
\end{equation} where $\bar{\alpha}_{\rm s/p}$ is the steady-state amplitude of the coherent state in the supporting/primary mode obeying $|\bar{\alpha}_{\rm s/p}|^2=\bar{n}_\textrm{s/p}$.

In addition to the driven occupation given by Eq.~\eqref{eq:n_steady}, the resonator mode has a thermal population $1/\left\{\exp[\hbar\omega_{\rm s}/(k_{\rm B}T)]-1\right\}$. We neglect this as it is orders of magnitude smaller than the driven occupation in the temperature scales $T_{\rm N}, T_{\rm T,s}$ of the experiment~\cite{Masuda2018,Silveri17}, although a more complex sample capable of a precise temperature measurement is needed to fully rule this effect out.

\section{Derivation of the effective Lamb shift arising from multiphoton-assisted electron tunnelling events}
\label{sec:lamb}

Here, we derive the dynamic effective Lamb shift of the primary mode, Eq.~\eqref{eq:lamb_dyn}, which originates from the photon-assisted electron tunnelling at the two capacitively coupled superconductor--insulator--normal-metal tunnel junctions. We take into account also the tunnelling events induced by an rf drive on the supporting mode. See the main text for the definitions of the special terms. The derivation begins by following the approach of refs.~\cite{Silveri17,Silveri19,Ingold92} but to the best of our knowledge has not been presented before in the literature in this case of two bosonic modes coupled to the QCR such that one of the modes is traced out.

Let us denote by $|\eta\rangle=\left|Q_{\mathrm{N}}, m_{\rm p}, m_{\rm s}, \ell, k\right\rangle=\left|Q_{\mathrm{N}}, m_{\rm p}, m_{\rm s}\right\rangle\otimes\left|\ell, k\right\rangle$ the eigenstate of the combined unperturbed system composed of the electrical circuit degrees of freedom $\left|Q_{\mathrm{N}}, m_{\rm p}, m_{\rm s}\right\rangle$ and those of the quasiparticles in the normal metal and superconductor of the QCR $\left|\ell, k\right\rangle$. Owing to the pertubation intorduced by the tunnelling Hamiltonian $\hat{H}_\textrm{T}$, the energy-level shift $\hbar \delta_{\eta}$ of $\ket{\eta}$, where $\hbar$ is the reduced Planck constant, is given by the second-order time-independent perturbation theory as~\cite{Silveri19} 
\begin{equation} \label{eq:suppl_energy_level_shift}
    \hbar \delta_{\eta}=E_{\eta}-E_{\eta}^{0}=-\sum_{\eta^{\prime} \neq \eta} \frac{\left|\left\langle\eta^{\prime}\left|\hat{H}_{\mathrm{T}}\right| \eta\right\rangle\right|^{2}}{E_{\eta^{\prime}}-E_{\eta}}, 
\end{equation}{}
where the total energy is $E_{\eta}=E_{\mathrm{N}} Q_{\mathrm{N}}^{2}+\hbar\omega_{\mathrm{p}}^{0} m_{\rm p}+\hbar\omega_{\rm s} m_{\rm s}+\varepsilon_{\ell}+\epsilon_{k}$. Here, the integer $Q_{\rm N}$ is the eigenvalue of charge for the normal-metal QCR island, $\omega_{\rm p}^0$ is the bare angular frequency of the primary mode, $\omega_{\rm s}$ is the angular frequency of the supporting mode, and $m_{\rm p/s}$ denotes the number of photons in the primary or the supporting mode. Furthermore, the charging energy of the QCR island is $E_{\rm N}=e^2/(2C_{\rm N}^\prime)$, where $C_{\rm N}^\prime$ is the renormalized capacitance of the island provided by Eq.~\eqref{eq:CN}. The quasiparticles of the normal-metal and superconducting electrodes have energies $\varepsilon_\ell$ and~$\epsilon_k$, respectively, where $\ell$ and $k$ index these and the corresponding eigenstates throughout this document implicitly including the spin degree of freedom. The quasiparticle tunnelling events between the normal-metal and superconducting leads are described by the tunnelling Hamiltonian~\cite{Ingold92} 
\begin{equation}
    \hat{H}_{\mathrm{T}} =\sum\limits_{\ell k}\left(T_{\ell k} \hat{d}_{\ell}^\dagger \hat{c}_{k}\mathrm{e}^{-\mathrm{i} \frac{e}{\hbar} \hat{\phi}_{\mathrm{N}}}+T_{\ell k}^{*} \hat{d}_{\ell} \hat{c}_{k}^{\dagger} \mathrm{e}^{\mathrm{i} \frac{e}{\hbar} \hat{\phi}_{\mathrm{N}}}\right),
\end{equation}{}
where $T_{\ell k}$ is a tunnelling matrix element, the annihilation operators of the normal-metal and of the superconducting leads are denoted by $\hat{d}_{\ell}$ and $\hat{c}_{k}$, respectively,  $\hat{\phi}_{\rm N}$ is the node flux operator of the normal-metal island, and $e$ is the elementary charge. These events cause weak perturbations to the coupled electric circuit shown in Fig.~\ref{fig:circuit}, described by the core Hamiltonian which we represent using its eigenstate decomposition as 
\begin{align}
    \hat{H}=&\sum\limits_{Q_\mathrm{N}=-\infty}^{\infty} \sum\limits_{m_{\mathrm{p}},m_{\mathrm{s}}=0}^{\infty}\left(E_{\mathrm{N}} Q_{\mathrm{N}}^{2}+\hbar \omega_{\mathrm{p}}^{0} m_{\mathrm{p}}+\hbar \omega_{\mathrm{s}} m_{\mathrm{s}}\right) \nonumber \\* 
    &\times\ket{Q_{\mathrm{N}}, m_{\mathrm{p}}, m_{\mathrm{s}}}\bra{Q_{\mathrm{N}}, m_{\mathrm{p}}, m_{\mathrm{s}}}, 
\end{align}{}
where the eigenstates of the core Hamiltonian are obtained from the  eigenstates of the charge $\ket{Q_\textrm{N}}$ and from the Fock states $\ket{m_\textrm{p}}$ and $\ket{m_\textrm{s}}$ of the primary and supporting modes as $\left|Q_{\mathrm{N}}, m_{\mathrm{p}}, m_{\mathrm{s}}\right\rangle=\exp \left(-\mathrm{i} \alpha_{\rm p} Q_{\mathrm{N}} \frac{e}{h} \hat{\Phi}_{\rm p} -\mathrm{i} \alpha_{\rm s} Q_{\mathrm{N}} \frac{e}{h} \hat{\Phi}_{\rm s}\right)\left|Q_{\mathrm{N}}\right\rangle\otimes|m_{\rm p}\rangle\otimes|m_{\rm s}\rangle,$ where $\hat{\Phi}_{\rm p/s}$ are the flux operators and $\alpha_{\rm p/s}$ the capacitance ratios of the resonator modes as described by Eq.~\eqref{eq:capacitance_ratio}.

We use the transition matrix elements $M_{m m^\prime}^{\rm (p/s)}$ of the electric circuit given in Eq.~\eqref{eq:M_app} and expand equation~\eqref{eq:suppl_energy_level_shift} assuming that $E_\textrm{N}$ is the smallest relevant energy scale of the system $\left(E_{\rm N}\ll\Delta, \hbar\omega_{\mathrm{p}}^{0}, \hbar\omega_{\rm s} \text { and } k_{\mathrm{B}} T_{\mathrm{N}}\right)$. After tracing out the charge degree of freedom of the normal-metal and of the superconducting leads, we obtain an energy shift for the Fock state $\ket{m_\textrm{p}}$ corresponding to the transition $\ket{m_\textrm{s}}\to\ket{m_\textrm{s}'}$ in the supporting mode as 
\begin{align}
    \hbar \tilde{\delta}&_{m_{\rm p},m_\textrm{s},m_\textrm{s}'} \nonumber \\*
    =&-\sum_{m_{\rm p}^{\prime}} \sum_{\ell, \ell^\prime} \sum_{k, k^\prime}\left|M_{m_{\rm p} m_{\rm p}^\prime}^{\rm (p)}\right|^{2}\left|M_{m_{\rm s} m_{\rm s}^\prime}^{\rm (s)}\right|^{2} \nonumber \\*
    &\left[ \frac{\left|\left\langle\ell^{\prime} k^\prime\left|\hat{\Theta}^{\dagger}\right| \ell k\right\rangle\right|^{2}}{E_{\mathrm{N}}+\ell_{\rm p}\hbar \omega_{\mathrm{p}}^{0}+\ell_{\rm s}\hbar \omega_{\rm s}+E_{\ell^\prime k^{\prime}}-E_{\ell k}}\right. \nonumber \\*
    &\quad+\left.\frac{\left|\left\langle\ell^{\prime} k^{\prime}|\hat{\Theta}| \ell k\right\rangle\right|^{2}}{E_{\mathrm{N}}+\ell_{\rm p}\hbar \omega_{\mathrm{p}}^{0}+\ell_{\rm s}\hbar \omega_{\rm s}+E_{\ell^\prime k^{\prime}}-E_{\ell k}} \right], 
\end{align}
where $\hat{\Theta}=\sum_{\ell k}T_{\ell k}\hat{d}_{\ell}^\dagger\hat{c}_{k}$ and its Hermitian conjugate is the quasiparticle part of the tunnelling Hamiltonian and $\ell_{\rm p/s} = m_{\rm p/s}^\prime-m_{\rm p/s}$ is the change in the mode occupations.

The interaction parameter of the primary mode is small $\rho_{\rm p}\approx0.003$, which allows us to expand $\left|M_{m_{\rm p}m_{\rm p}^\prime}^{\rm (p)}\right|^{2}$ up to the first order in $\rho_{\rm p}$. The corresponding dynamic effective Lamb shift of the primary mode $\omega_{\mathrm{L,p},m_\textrm{s},m_\textrm{s}'}=\tilde{\delta}_{m_{\rm p}+1,m_\textrm{s},m_\textrm{s}'}-\tilde{\delta}_{m_{\rm p},m_\textrm{s},m_\textrm{s}'}$ is thus given by 
\begin{align} \label{eq:suppl_dyn_lamb_shift}
    \omega&_{\mathrm{L,p},m_\textrm{s},m_\textrm{s}'} \nonumber \\*
    =&\frac{\rho_{\rm p}}{\hbar} \sum_{\ell, \ell^\prime} \sum_{k, k^\prime} \sum_{\ell_{\rm p}=\pm 1}\left|M_{m_{\rm s} m_{\rm s}^\prime}^{\rm (s)}\right|^{2} \nonumber \\*
    &\left[\frac{\left|\left\langle\ell^{\prime} k^{\prime}\left|\hat{\Theta}^{\dagger}\right| \ell k\right\rangle\right|^{2}+\left|\left\langle\ell^{\prime} k^\prime|\hat{\Theta}| \ell k\right\rangle\right|^{2}}{E_{\mathrm{N}}+\ell_{\rm s}\hbar\omega_{\rm s}+E_{\ell^\prime k^\prime}- E_{\ell k}}\right. \nonumber \\*
    &\quad-\left.\frac{\left|\left\langle\ell^{\prime} k^{\prime}\left|\hat{\Theta}^{\dagger}\right| \ell k\right\rangle\right|^{2}+\left|\left\langle\ell^{\prime} k^\prime|\hat{\Theta}| \ell k\right\rangle\right|^{2}}{E_{\mathrm{N}}+\ell_{\rm p}\hbar\omega_{\rm p}^0+\ell_{\rm s}\hbar\omega_{\rm s}+E_{\ell^\prime k^\prime}- E_{\ell k}}\right].
\end{align}
Here, we denote the first term owing to elastic transitions as $\omega_{{\rm L,p},m_\textrm{s},m_\textrm{s}'}^{\rm el}$ and the second term owing to the primary-mode-assisted transitions as $\omega_{{\rm L,p},m_\textrm{s},m_\textrm{s}'}^{\rm ph}$. Note that the elastic transition term is simply $\omega_{{\rm L,p},m_\textrm{s},m_\textrm{s}'}^{\rm el} =-\lim _{\omega_{\rm p}^0 \rightarrow 0} \omega_{{\rm L,p},m_\textrm{s},m_\textrm{s}'}^{\rm ph}\left(\omega_{\mathrm{p}}^{0}\right)$ and it thus suffices to simplify the primary-mode-assisted part.

We express the matrix elements of the quasiparticle transitions $\left|\left\langle\ell^{\prime} k^{\prime}|\hat{\Theta}| \ell k\right\rangle\right|$ with the normalized forward quasiparticle tunnelling rate $F$ given in Eq.~\eqref{eq:F}. Here, we assume that the tunnelling matrix elements are approximately constant in the vicinity of the Fermi energies, the two SIN junctions are identical, and the electrodes are at equal temperature. Consequently, we obtain
\begin{align} \label{eq:suppl_dyn_ph_lamb_shift_sum_simple}
    \omega&_{\mathrm{L,p},m_\textrm{s},m_\textrm{s}'}^{\mathrm{ph}} \nonumber \\*
    =&-\frac{\rho_{\rm p}}{\pi}\frac{R_{\mathrm{K}}}{R_{\mathrm{T}}}\left|M_{m_{\rm s} m_{\rm s}^\prime}^{\rm (s)}\right|^{2} \sum_{\tau,\ell_{\rm p}=\pm 1} \mathrm{PV} \int_{-\infty}^{\infty} \mathrm{d} \epsilon \nonumber \\*
    &\frac{F\left(\epsilon+\tau e V+\ell_{\rm p} \hbar \omega_{\mathrm{p}}^{0}+\ell_{\rm s}\hbar\omega_{\rm s}-E_{\mathrm{N}}\right)}{\epsilon} \nonumber \\*
    =&-2\pi \alpha_{\rm p}^{2} \frac{Z_{\rm p}}{R_{\rm T}}\left|M_{m_{\rm s} m_{\rm s}^\prime}^{\rm (s)}\right|^{2} \sum_{\tau, \ell_{\rm p}=\pm 1}  \left[ \mathrm{PV} \int_{0}^{\infty} \frac{\mathrm{d}\omega}{2 \pi}\right. \nonumber \\*
    &\frac{ \ell_{\rm p} F\left(\tau e V+\ell_{\rm p} \hbar \omega+\ell_{\rm s}\hbar\omega_{\rm s}-E_{N}\right)}{\omega -\omega_{\mathrm{p}}^{0}}+\mathrm{PV} \nonumber \\*
    &\left. \int_{0}^{\infty} \frac{\mathrm{d} \omega}{2 \pi} \frac{\ell_{\rm p} F\left(\tau e V+\ell_{\rm p} \hbar \omega+\ell_{\rm s}\hbar\omega_{\rm s}-E_{\mathrm{N}}\right)}{\omega+\omega_{\mathrm{p}}^{0}}\right], 
\end{align}
where PV denotes the Cauchy principal value integration, $R_\textrm{K}$ is the von Klitzing constant, $R_\textrm{T}$ is the tunnelling resistance, and $Z_\textrm{p}$ is the characteristic impedance of the primary mode in its lumped-element description. Furthermore, by tracing out the supporting resonator mode with occupation probabilities $P_{m_\textrm{s}}$ and expressing equation~\eqref{eq:suppl_dyn_ph_lamb_shift_sum_simple} in terms of the coupling strength $\gamma_{\mathrm{T,p}}$ of the effective electromagnetic environment 
\begin{align}
    \gamma_{\mathrm{T,p}}(V, \omega_{\rm p})=&2\pi \alpha_{\rm p}^{2} \frac{Z_{\mathrm{p}}}{R_{\rm T}}\sum_{m_\textrm{s},m_\textrm{s}'}P_{m_\textrm{s}}\left|M_{m_\textrm{s}m_\textrm{s}'}^{\rm (s)}\right|^{2} \sum_{\tau,\ell_{\rm p}=\pm 1} \nonumber \\*
    &\ell_{\rm p} F\left(\tau e V+\ell_{\rm p} \hbar \omega_{\rm p}+\ell_{\rm s}\hbar\omega_{\rm s}-E_{\mathrm{N}}\right), 
\end{align}{}
we obtain the dynamic effective Lamb shift of the primary mode 
\begin{align}
    \omega_{\mathrm{L,p}}\left(V, \omega_{\mathrm{p}}^{0}\right)=-\mathrm{PV} \int_{0}^{\infty} \frac{\mathrm{d} \omega}{2 \pi}&\left[\frac{\gamma_{\mathrm{T,p}}(V, \omega)}{\omega-\omega_{\mathrm{p}}^{0}}+\frac{\gamma_{\mathrm{T,p}}(V, \omega)}{\omega+\omega_{\mathrm{p}}^{0}}\right. \nonumber \\*
    &\left.-2 \frac{\gamma_{\mathrm{T,p}}(V, \omega)}{\omega}\right], 
\end{align}{}
where the terms dependent on the primary mode frequency $\omega_{\rm p}^0$ are contributions from the photon-assisted transitions and the last term arises from elastic tunnelling. Note that the effective Lamb shift of the primary mode depends on the quantum state of the supporting mode through the coupling strength $\gamma_{\rm T,p}$ and the occupation probabilities $P_{m_\textrm{s}}$ of the Fock states $\ket{m_\textrm{s}}$.

\section{Theory of the Lamb shift of a bosonic mode coupled to a fermionic bath} \label{sec:lamb_general}

\begin{figure}
    \centering
    \includegraphics[width=\columnwidth]{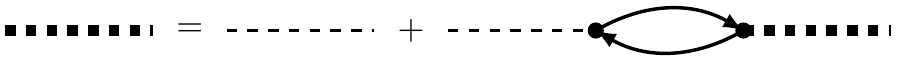}
    \caption{Diagrammatic form of the Dyson equation for the bosonic Green's function. Here, the thin dashed lines correspond to the bare Green's function of the bosonic mode, the thin solid lines correspond to the bare Green's functions of the fermionic bath, and the thick dashed lines correspond to the Green's function of the bosonic mode with the interactions taken into account. The considered diagrams correspond to the Born approximation for the bosonic Green's function.}
    \label{fig:diagram}
\end{figure}

We study a primary bosonic mode coupled to a bath of non-interacting fermions described by the following Hamiltonian:
\begin{align}
    \hat H =& \hbar \omega_{\mathrm p}^0 \hat b^\dag \hat b + \sum\limits_k \xi_k \hat a_k^\dag
    \hat a_k \nonumber \\*
    &+ \sum\limits_{kk'} \left(\gamma_{kk'} \hat b \hat a_k^\dag \hat
    a_{k'} + \gamma_{kk'}^\ast \hat b^\dag \hat a_{k'}^\dag \hat a_k\right),
    \label{eq:bf-hamiltonian}
\end{align}
where $\omega_{\mathrm p}^0$ is the bare frequency of the primary mode, $\xi_k$ is the energy of $k$-th fermionic mode of the bath, and the parameters $\gamma_{kk'}$ control coupling between the bosonic mode and the fermionic bath. We do not consider excitations of the bath arising from any coupling it to a driven supporting mode. Instead we assume that the whole system has some equilibrium but nonzero temperature $T$ and study the temperature dependence of the primary-mode frequency. This assumption allows us to use a finite-temperature diagrammatic technique for this problem. 

We begin by defining an imaginary-time bosonic Green's function $D(\tau, \tau')$ as 
\begin{equation}
    D(\tau, \tau') = \frac{\mathop{\mathrm{Tr}} \left\{ e^{-\beta \hat H} \mathbb T_\tau \left[
        \hat b(\tau) \hat b^\dag(\tau')\right]\right\}}{\mathop{\mathrm{Tr}} e^{-\beta \hat
    H}},
\end{equation}
where $\beta = (k_\textrm{B} T)^{-1}$,  $\mathbb T_\tau$ is the imaginary-time-ordering operator and the operators $\hat b(\tau)$ and $\hat b^\dag(\tau)$ are defined as
\begin{gather}
    \hat b(\tau) = e^{\frac{\hat H \tau}{\hbar}} \hat b e^{-\frac{\hat H \tau}{\hbar}}, \\
    \hat b^\dag(\tau) = e^{\frac{\hat H \tau}{\hbar}} \hat b^\dag e^{-\frac{\hat H \tau}{\hbar}}.
\end{gather}
This Green's function is defined in the interval $-{\hbar}/(k_\textrm{B} T) < \tau - \tau' < {\hbar}/(k_\textrm{B} T)$ and can be expanded into a Fourier series as
\begin{equation}
    D(\tau, \tau') = \frac{k_\textrm{B} T}{\hbar} \sum\limits_{n=-\infty}^{\infty} D(\omega_n^\textrm{b}) e^{-i (\tau -
    \tau')\omega_n^\textrm{b}},
\end{equation}
where $\omega_n^\textrm{b} = 2\pi n k_\textrm{B} T / \hbar$ are bosonic Matsubara frequencies.
For the single bosonic mode uncoupled from the bath, the Fourier expansion coefficients are equal to $D^{(0)}(\omega_n^\textrm{b}) = (i \omega_n^\textrm{b} - \omega_\textrm{p}^0)^{-1}$. Thus, if we consider $i \omega_n^\textrm{b}$ as a complex variable, there is a pole at the oscillator frequency $i \omega_n^\textrm{b} = \omega_\textrm{p}^0$. This property allows us to define the bath-shifted frequency of the mode as the position of the pole of the Green's function of the interacting system. The Green's function of the interacting system can be expanded as a series of
diagrams consisting of the Green's functions of the non-interacting (bare) bosonic and fermionic modes. The main series of contributions which corresponds to the Born approximation for the Green's function is shown in the Fig.~\ref{fig:diagram}. The Dyson equation corresponding to these diagrams reads as follows:
\begin{gather}\label{eq:dyson}
    \left[D\left(\omega_n^\textrm{b}\right)\right]^{-1} =
    \left[D^{(0)}\left(\omega_n^\textrm{b}\right)\right]^{-1} - \Pi(\omega_n^\textrm{b}), \\
    \label{eq:dyson2}
    \Pi(\omega_n^\textrm{b}) = \frac{k_\textrm{B} T}{\hbar^3} \sum\limits_{kk'}  \sum\limits_{m=-\infty}^{+\infty} \frac{|\gamma_{kk'}|^2}{\left(i \omega_m^\textrm{f} + i \omega_n^\textrm{b} - \xi_k\right)\left(i \omega_m^\textrm{f} - \xi_{k'}\right)},
\end{gather}
where $\omega_m^\textrm{f} = \pi (2 m + 1) k_\textrm{B} T / \hbar$ are the fermionic Matsubara frequencies. Vanishing right side of the Dyson equation~\eqref{eq:dyson} correspond to the pole of the Green's function, and, thus, its position yields the bath-shifted frequency.

To proceed, let us introduce simplifying assumptions on the structure of the bath and the coupling parameters. We assume that the energies of the fermionic modes of the bath are uniformly distributed within the interval $-\mu$ to $W - \mu$, where $\mu$ is the chemical potential and $W$ is the width of the bath. Thus, we can replace summation over $k$ with an integral $\sum_k \to \nu V_\textrm{bath} \int_{-\mu}^{W - \mu}\;\mathrm{d}\xi$, where $\nu$ is the density of states per the unit volume and $V_\textrm{bath}$ is the volume of the Fermi gas. In addition, we assume that absorption and emission of a boson couples all fermionic modes between each other and fix $|\gamma_{kk'}|^2 = \Gamma^2 / V_\textrm{bath}^2$. Consequently, we have a single parameter which controls the coupling strength. Under these assumptions the polarization operator $\Pi(\omega_n^\textrm{b})$ can be evaluated as
\begin{align}
    \Pi(\omega_n^\textrm{b}) = \frac{\Gamma^2 \nu^2}{2\hbar} \int\limits_{-\mu}^{W - \mu} \ln &\left[ 
    \frac{(\varepsilon - W + \mu)^2 + (\hbar \omega_n^\textrm{b})^2}{(\varepsilon+\mu)^2 + (\hbar \omega_n^\textrm{b})^2}
    \right] \nonumber \\*
    &\tanh \frac{\varepsilon}{2 k_\textrm{B} T}\;\mathrm{d}\varepsilon .
\end{align}
Finally, the equation for the frequency shift can be expressed as
\begin{align}
    \omega_\mathrm{L,p}(T) =& \frac{\Gamma^2 \nu^2}{2\hbar} \int\limits_{-\mu}^{W - \mu} \nonumber \\*
    &\ln \left\{ 
    \frac{(\varepsilon - W + \mu)^2 + [\hbar (\omega_\mathrm{L,p} + \omega_\mathrm{p}^0) + i 0]^2}{(\varepsilon+\mu)^2 - [\hbar (\omega_\mathrm{L,p} + \omega_\mathrm{p}^0) + i 0]^2}
    \right\}\nonumber \\*
    &\tanh \frac{\varepsilon}{2 k_\textrm{B} T}\;\mathrm{d}\varepsilon .
\end{align}
In the limit $\Gamma \nu \ll 1$ and $\hbar \omega_\textrm{p}^0, k_\textrm{B} T \ll \mu, (W - \mu)$ the approximate solution of this equation can be found as
\begin{align}
    \omega_\mathrm{L,p}(T) \approx& \frac{\Gamma^2 \nu^2}{\hbar} \int\limits_{-\mu}^{W- \mu}\ln \left(\frac{W - \mu + \varepsilon}{\varepsilon + \mu}\right)\tanh \frac{\varepsilon}{k_\textrm{B} T}\; \mathrm{d}\varepsilon \nonumber \\*
    &- i \pi \Gamma^2 \nu^2 \omega_\mathrm{p}^0 \nonumber \\*
    \approx& \omega_\mathrm{L,p}(0) - \frac{\pi^2 \Gamma^2 \nu^2 (k_\textrm{B} T)^2 W}{3 \hbar \mu (W - \mu)} \nonumber \\*
    &- i \pi \Gamma^2 \nu^2 \omega_\mathrm{p}^0.
\end{align}
The real part of $\omega_\mathrm{L,p}$ defines the bath-induced frequency shift of the bosonic mode whereas the imaginary part gives the dissipation rate.


\begin{thebibliography}{40}%
\makeatletter
\providecommand \@ifxundefined [1]{%
 \@ifx{#1\undefined}
}%
\providecommand \@ifnum [1]{%
 \ifnum #1\expandafter \@firstoftwo
 \else \expandafter \@secondoftwo
 \fi
}%
\providecommand \@ifx [1]{%
 \ifx #1\expandafter \@firstoftwo
 \else \expandafter \@secondoftwo
 \fi
}%
\providecommand \natexlab [1]{#1}%
\providecommand \enquote  [1]{``#1''}%
\providecommand \bibnamefont  [1]{#1}%
\providecommand \bibfnamefont [1]{#1}%
\providecommand \citenamefont [1]{#1}%
\providecommand \href@noop [0]{\@secondoftwo}%
\providecommand \href [0]{\begingroup \@sanitize@url \@href}%
\providecommand \@href[1]{\@@startlink{#1}\@@href}%
\providecommand \@@href[1]{\endgroup#1\@@endlink}%
\providecommand \@sanitize@url [0]{\catcode `\\12\catcode `\$12\catcode
  `\&12\catcode `\#12\catcode `\^12\catcode `\_12\catcode `\%12\relax}%
\providecommand \@@startlink[1]{}%
\providecommand \@@endlink[0]{}%
\providecommand \url  [0]{\begingroup\@sanitize@url \@url }%
\providecommand \@url [1]{\endgroup\@href {#1}{\urlprefix }}%
\providecommand \urlprefix  [0]{URL }%
\providecommand \Eprint [0]{\href }%
\providecommand \doibase [0]{https://doi.org/}%
\providecommand \selectlanguage [0]{\@gobble}%
\providecommand \bibinfo  [0]{\@secondoftwo}%
\providecommand \bibfield  [0]{\@secondoftwo}%
\providecommand \translation [1]{[#1]}%
\providecommand \BibitemOpen [0]{}%
\providecommand \bibitemStop [0]{}%
\providecommand \bibitemNoStop [0]{.\EOS\space}%
\providecommand \EOS [0]{\spacefactor3000\relax}%
\providecommand \BibitemShut  [1]{\csname bibitem#1\endcsname}%
\let\auto@bib@innerbib\@empty
\bibitem [{\citenamefont {Ernst}\ \emph {et~al.}(1987)\citenamefont {Ernst},
  \citenamefont {Bodenhausen},\ and\ \citenamefont {Wokaun}}]{Ernst1987}%
  \BibitemOpen
  \bibfield  {author} {\bibinfo {author} {\bibfnamefont {R.~R.}\ \bibnamefont
  {Ernst}}, \bibinfo {author} {\bibfnamefont {G.}~\bibnamefont {Bodenhausen}},\
  and\ \bibinfo {author} {\bibfnamefont {A.}~\bibnamefont {Wokaun}},\
  }\href@noop {} {\emph {\bibinfo {title} {Principles of Nuclear Magnetic
  Resonance in One and Two Dimensions}}}\ (\bibinfo  {publisher} {Clarendon
  press},\ \bibinfo {address} {Oxford},\ \bibinfo {year} {1987})\BibitemShut
  {NoStop}%
\bibitem [{\citenamefont {H{\"a}m{\"a}l{\"a}inen}\ \emph
  {et~al.}(1993)\citenamefont {H{\"a}m{\"a}l{\"a}inen}, \citenamefont {Hari},
  \citenamefont {Ilmoniemi}, \citenamefont {Knuutila},\ and\ \citenamefont
  {Lounasmaa}}]{Hamalainen1993}%
  \BibitemOpen
  \bibfield  {author} {\bibinfo {author} {\bibfnamefont {M.}~\bibnamefont
  {H{\"a}m{\"a}l{\"a}inen}}, \bibinfo {author} {\bibfnamefont {R.}~\bibnamefont
  {Hari}}, \bibinfo {author} {\bibfnamefont {R.~J.}\ \bibnamefont {Ilmoniemi}},
  \bibinfo {author} {\bibfnamefont {J.}~\bibnamefont {Knuutila}},\ and\
  \bibinfo {author} {\bibfnamefont {O.~V.}\ \bibnamefont {Lounasmaa}},\
  }\bibfield  {title} {\bibinfo {title} {Magnetoencephalography—theory,
  instrumentation, and applications to noninvasive studies of the working human
  brain},\ }\href {https://doi.org/10.1103/RevModPhys.65.413} {\bibfield
  {journal} {\bibinfo  {journal} {Rev. Mod. Phys.}\ }\textbf {\bibinfo {volume}
  {65}},\ \bibinfo {pages} {413} (\bibinfo {year} {1993})}\BibitemShut
  {NoStop}%
\bibitem [{\citenamefont {Ladd}\ \emph {et~al.}(2010)\citenamefont {Ladd},
  \citenamefont {Jelezko}, \citenamefont {Laflamme}, \citenamefont {Nakamura},
  \citenamefont {Monroe},\ and\ \citenamefont {O’Brien}}]{Ladd10}%
  \BibitemOpen
  \bibfield  {author} {\bibinfo {author} {\bibfnamefont {T.~D.}\ \bibnamefont
  {Ladd}}, \bibinfo {author} {\bibfnamefont {F.}~\bibnamefont {Jelezko}},
  \bibinfo {author} {\bibfnamefont {R.}~\bibnamefont {Laflamme}}, \bibinfo
  {author} {\bibfnamefont {Y.}~\bibnamefont {Nakamura}}, \bibinfo {author}
  {\bibfnamefont {C.}~\bibnamefont {Monroe}},\ and\ \bibinfo {author}
  {\bibfnamefont {J.~L.}\ \bibnamefont {O’Brien}},\ }\bibfield  {title}
  {\bibinfo {title} {Quantum computers},\ }\href
  {https://doi.org/10.1038/nature08812} {\bibfield  {journal} {\bibinfo
  {journal} {Nature}\ }\textbf {\bibinfo {volume} {464}},\ \bibinfo {pages}
  {45} (\bibinfo {year} {2010})}\BibitemShut {NoStop}%
\bibitem [{\citenamefont {Neill}\ \emph {et~al.}(2018)\citenamefont {Neill},
  \citenamefont {Roushan}, \citenamefont {Kechedzhi}, \citenamefont {Boixo},
  \citenamefont {Isakov}, \citenamefont {Smelyanskiy}, \citenamefont {Megrant},
  \citenamefont {Chiaro}, \citenamefont {Dunsworth},\ and\ \citenamefont
  {Arya~et al.}}]{Neill18}%
  \BibitemOpen
  \bibfield  {author} {\bibinfo {author} {\bibfnamefont {C.}~\bibnamefont
  {Neill}}, \bibinfo {author} {\bibfnamefont {P.}~\bibnamefont {Roushan}},
  \bibinfo {author} {\bibfnamefont {K.}~\bibnamefont {Kechedzhi}}, \bibinfo
  {author} {\bibfnamefont {S.}~\bibnamefont {Boixo}}, \bibinfo {author}
  {\bibfnamefont {S.~V.}\ \bibnamefont {Isakov}}, \bibinfo {author}
  {\bibfnamefont {V.}~\bibnamefont {Smelyanskiy}}, \bibinfo {author}
  {\bibfnamefont {A.}~\bibnamefont {Megrant}}, \bibinfo {author} {\bibfnamefont
  {B.}~\bibnamefont {Chiaro}}, \bibinfo {author} {\bibfnamefont
  {A.}~\bibnamefont {Dunsworth}},\ and\ \bibinfo {author} {\bibfnamefont
  {K.}~\bibnamefont {Arya~et al.}},\ }\bibfield  {title} {\bibinfo {title} {A
  blueprint for demonstrating quantum supremacy with superconducting qubits},\
  }\href {https://doi.org/10.1126/science.aao4309} {\bibfield  {journal}
  {\bibinfo  {journal} {Science}\ }\textbf {\bibinfo {volume} {360}},\ \bibinfo
  {pages} {195} (\bibinfo {year} {2018})}\BibitemShut {NoStop}%
\bibitem [{\citenamefont {Rosenblum}\ \emph {et~al.}(2018)\citenamefont
  {Rosenblum}, \citenamefont {Reinhold}, \citenamefont {Mirrahimi},
  \citenamefont {Jiang}, \citenamefont {Frunzio},\ and\ \citenamefont
  {Schoelkopf}}]{Rosenblum18}%
  \BibitemOpen
  \bibfield  {author} {\bibinfo {author} {\bibfnamefont {S.}~\bibnamefont
  {Rosenblum}}, \bibinfo {author} {\bibfnamefont {P.}~\bibnamefont {Reinhold}},
  \bibinfo {author} {\bibfnamefont {M.}~\bibnamefont {Mirrahimi}}, \bibinfo
  {author} {\bibfnamefont {L.}~\bibnamefont {Jiang}}, \bibinfo {author}
  {\bibfnamefont {L.}~\bibnamefont {Frunzio}},\ and\ \bibinfo {author}
  {\bibfnamefont {R.~J.}\ \bibnamefont {Schoelkopf}},\ }\bibfield  {title}
  {\bibinfo {title} {Fault-tolerant detection of a quantum error},\ }\href
  {https://doi.org/10.1126/science.aat3996} {\bibfield  {journal} {\bibinfo
  {journal} {Science}\ }\textbf {\bibinfo {volume} {361}},\ \bibinfo {pages}
  {266} (\bibinfo {year} {2018})}\BibitemShut {NoStop}%
\bibitem [{\citenamefont {Arute}\ \emph {et~al.}(2019)\citenamefont {Arute},
  \citenamefont {Arya}, \citenamefont {Babbush}, \citenamefont {Bacon},
  \citenamefont {Bardin}, \citenamefont {Barends}, \citenamefont {Biswas},
  \citenamefont {Boixo}, \citenamefont {Brandao},\ and\ \citenamefont {{Buell
  et al.}}}]{Arute2019}%
  \BibitemOpen
  \bibfield  {author} {\bibinfo {author} {\bibfnamefont {F.}~\bibnamefont
  {Arute}}, \bibinfo {author} {\bibfnamefont {K.}~\bibnamefont {Arya}},
  \bibinfo {author} {\bibfnamefont {R.}~\bibnamefont {Babbush}}, \bibinfo
  {author} {\bibfnamefont {D.}~\bibnamefont {Bacon}}, \bibinfo {author}
  {\bibfnamefont {J.}~\bibnamefont {Bardin}}, \bibinfo {author} {\bibfnamefont
  {R.}~\bibnamefont {Barends}}, \bibinfo {author} {\bibfnamefont
  {R.}~\bibnamefont {Biswas}}, \bibinfo {author} {\bibfnamefont
  {S.}~\bibnamefont {Boixo}}, \bibinfo {author} {\bibfnamefont
  {F.}~\bibnamefont {Brandao}},\ and\ \bibinfo {author} {\bibfnamefont
  {D.}~\bibnamefont {{Buell et al.}}},\ }\bibfield  {title} {\bibinfo {title}
  {Quantum supremacy using a programmable superconducting processor},\ }\href
  {https://doi.org/10.1038/s41586-019-1666-5} {\bibfield  {journal} {\bibinfo
  {journal} {Nature}\ }\textbf {\bibinfo {volume} {574}},\ \bibinfo {pages}
  {505} (\bibinfo {year} {2019})}\BibitemShut {NoStop}%
\bibitem [{\citenamefont {Blais}\ \emph {et~al.}(2020)\citenamefont {Blais},
  \citenamefont {Girvin},\ and\ \citenamefont {Oliver}}]{Blais20}%
  \BibitemOpen
  \bibfield  {author} {\bibinfo {author} {\bibfnamefont {A.}~\bibnamefont
  {Blais}}, \bibinfo {author} {\bibfnamefont {S.~M.}\ \bibnamefont {Girvin}},\
  and\ \bibinfo {author} {\bibfnamefont {W.~D.}\ \bibnamefont {Oliver}},\
  }\bibfield  {title} {\bibinfo {title} {Quantum information processing and
  quantum optics with circuit quantum electrodynamics},\ }\href
  {https://doi.org/10.1038/s41567-020-0806-z} {\bibfield  {journal} {\bibinfo
  {journal} {Nat. Phys.}\ }\textbf {\bibinfo {volume} {16}},\ \bibinfo {pages}
  {247} (\bibinfo {year} {2020})}\BibitemShut {NoStop}%
\bibitem [{\citenamefont {Bennett}\ and\ \citenamefont
  {Brassard}(1984)}]{Bennett84}%
  \BibitemOpen
  \bibfield  {author} {\bibinfo {author} {\bibfnamefont {C.~H.}\ \bibnamefont
  {Bennett}}\ and\ \bibinfo {author} {\bibfnamefont {G.}~\bibnamefont
  {Brassard}},\ }\bibfield  {title} {\bibinfo {title} {Quantum cryptography:
  Public key distribution and coin tossing},\ }in\ \href
  {https://arxiv.org/abs/2003.06557} {\emph {\bibinfo {booktitle} {Proc. IEEE
  Int. Conf. on Computers, Systems, and Signal Processing, Bangalore}}}\
  (\bibinfo {year} {1984})\ pp.\ \bibinfo {pages} {175--179}\BibitemShut
  {NoStop}%
\bibitem [{\citenamefont {Ekert}(1991)}]{Ekert91}%
  \BibitemOpen
  \bibfield  {author} {\bibinfo {author} {\bibfnamefont {A.~K.}\ \bibnamefont
  {Ekert}},\ }\bibfield  {title} {\bibinfo {title} {Quantum cryptography based
  on {B}ell's theorem},\ }\href {https://doi.org/10.1103/PhysRevLett.67.661}
  {\bibfield  {journal} {\bibinfo  {journal} {Phys. Rev. Lett.}\ }\textbf
  {\bibinfo {volume} {67}},\ \bibinfo {pages} {661} (\bibinfo {year}
  {1991})}\BibitemShut {NoStop}%
\bibitem [{\citenamefont {Lamb}\ and\ \citenamefont
  {Retherford}(1947)}]{Lamb47}%
  \BibitemOpen
  \bibfield  {author} {\bibinfo {author} {\bibfnamefont {W.~E.}\ \bibnamefont
  {Lamb}}\ and\ \bibinfo {author} {\bibfnamefont {R.~C.}\ \bibnamefont
  {Retherford}},\ }\bibfield  {title} {\bibinfo {title} {Fine {structure} of
  the {hydrogen} {atom} by a {microwave} {method}},\ }\href
  {https://doi.org/10.1103/PhysRev.72.241} {\bibfield  {journal} {\bibinfo
  {journal} {Phys. Rev.}\ }\textbf {\bibinfo {volume} {72}},\ \bibinfo {pages}
  {241} (\bibinfo {year} {1947})}\BibitemShut {NoStop}%
\bibitem [{\citenamefont {Bethe}(1947)}]{Bethe47}%
  \BibitemOpen
  \bibfield  {author} {\bibinfo {author} {\bibfnamefont {H.~A.}\ \bibnamefont
  {Bethe}},\ }\bibfield  {title} {\bibinfo {title} {The electromagnetic shift
  of energy levels},\ }\href {https://doi.org/10.1103/PhysRev.72.339}
  {\bibfield  {journal} {\bibinfo  {journal} {Phys. Rev.}\ }\textbf {\bibinfo
  {volume} {72}},\ \bibinfo {pages} {339} (\bibinfo {year} {1947})}\BibitemShut
  {NoStop}%
\bibitem [{\citenamefont {Bardeen}\ \emph {et~al.}(1957)\citenamefont
  {Bardeen}, \citenamefont {Cooper},\ and\ \citenamefont
  {Schrieffer}}]{Bardeen57}%
  \BibitemOpen
  \bibfield  {author} {\bibinfo {author} {\bibfnamefont {J.}~\bibnamefont
  {Bardeen}}, \bibinfo {author} {\bibfnamefont {L.~N.}\ \bibnamefont
  {Cooper}},\ and\ \bibinfo {author} {\bibfnamefont {J.~R.}\ \bibnamefont
  {Schrieffer}},\ }\bibfield  {title} {\bibinfo {title} {Theory of
  superconductivity},\ }\href {https://doi.org/10.1103/PhysRev.108.1175}
  {\bibfield  {journal} {\bibinfo  {journal} {Phys. Rev.}\ }\textbf {\bibinfo
  {volume} {108}},\ \bibinfo {pages} {1175} (\bibinfo {year}
  {1957})}\BibitemShut {NoStop}%
\bibitem [{\citenamefont {Casimir}\ and\ \citenamefont
  {Polder}(1948)}]{Casimir48}%
  \BibitemOpen
  \bibfield  {author} {\bibinfo {author} {\bibfnamefont {H.~B.~G.}\
  \bibnamefont {Casimir}}\ and\ \bibinfo {author} {\bibfnamefont
  {D.}~\bibnamefont {Polder}},\ }\bibfield  {title} {\bibinfo {title} {The
  influence of retardation on the {L}ondon-van der {W}aals forces},\ }\href
  {https://doi.org/10.1103/PhysRev.73.360} {\bibfield  {journal} {\bibinfo
  {journal} {Phys. Rev.}\ }\textbf {\bibinfo {volume} {73}},\ \bibinfo {pages}
  {360} (\bibinfo {year} {1948})}\BibitemShut {NoStop}%
\bibitem [{\citenamefont {Giaever}(1960)}]{Giaever60}%
  \BibitemOpen
  \bibfield  {author} {\bibinfo {author} {\bibfnamefont {I.}~\bibnamefont
  {Giaever}},\ }\bibfield  {title} {\bibinfo {title} {Energy gap in
  superconductors measured by electron tunneling},\ }\href
  {https://doi.org/10.1103/PhysRevLett.5.147} {\bibfield  {journal} {\bibinfo
  {journal} {Phys. Rev. Lett.}\ }\textbf {\bibinfo {volume} {5}},\ \bibinfo
  {pages} {147} (\bibinfo {year} {1960})}\BibitemShut {NoStop}%
\bibitem [{\citenamefont {Heinzen}\ and\ \citenamefont
  {Feld}(1987)}]{Heinzen87}%
  \BibitemOpen
  \bibfield  {author} {\bibinfo {author} {\bibfnamefont {D.~J.}\ \bibnamefont
  {Heinzen}}\ and\ \bibinfo {author} {\bibfnamefont {M.~S.}\ \bibnamefont
  {Feld}},\ }\bibfield  {title} {\bibinfo {title} {Vacuum {radiative} {level}
  {shift} and {spontaneous}-{emission} {linewidth} of an {atom} in an {optical}
  {resonator}},\ }\href {https://doi.org/10.1103/PhysRevLett.59.2623}
  {\bibfield  {journal} {\bibinfo  {journal} {Phys. Rev. Lett.}\ }\textbf
  {\bibinfo {volume} {59}},\ \bibinfo {pages} {2623} (\bibinfo {year}
  {1987})}\BibitemShut {NoStop}%
\bibitem [{\citenamefont {Brune}\ \emph {et~al.}(1994)\citenamefont {Brune},
  \citenamefont {Nussenzveig}, \citenamefont {Schmidt-Kaler}, \citenamefont
  {Bernardot}, \citenamefont {Maali}, \citenamefont {Raimond},\ and\
  \citenamefont {Haroche}}]{Brune94}%
  \BibitemOpen
  \bibfield  {author} {\bibinfo {author} {\bibfnamefont {M.}~\bibnamefont
  {Brune}}, \bibinfo {author} {\bibfnamefont {P.}~\bibnamefont {Nussenzveig}},
  \bibinfo {author} {\bibfnamefont {F.}~\bibnamefont {Schmidt-Kaler}}, \bibinfo
  {author} {\bibfnamefont {F.}~\bibnamefont {Bernardot}}, \bibinfo {author}
  {\bibfnamefont {A.}~\bibnamefont {Maali}}, \bibinfo {author} {\bibfnamefont
  {J.~M.}\ \bibnamefont {Raimond}},\ and\ \bibinfo {author} {\bibfnamefont
  {S.}~\bibnamefont {Haroche}},\ }\bibfield  {title} {\bibinfo {title} {From
  {L}amb shift to light shifts: {V}acuum and subphoton cavity fields measured
  by atomic phase sensitive detection},\ }\href
  {https://doi.org/10.1103/PhysRevLett.72.3339} {\bibfield  {journal} {\bibinfo
   {journal} {Phys. Rev. Lett.}\ }\textbf {\bibinfo {volume} {72}},\ \bibinfo
  {pages} {3339} (\bibinfo {year} {1994})}\BibitemShut {NoStop}%
\bibitem [{\citenamefont {Marrocco}\ \emph {et~al.}(1998)\citenamefont
  {Marrocco}, \citenamefont {Weidinger}, \citenamefont {Sang},\ and\
  \citenamefont {Walther}}]{Marrocco98}%
  \BibitemOpen
  \bibfield  {author} {\bibinfo {author} {\bibfnamefont {M.}~\bibnamefont
  {Marrocco}}, \bibinfo {author} {\bibfnamefont {M.}~\bibnamefont {Weidinger}},
  \bibinfo {author} {\bibfnamefont {R.~T.}\ \bibnamefont {Sang}},\ and\
  \bibinfo {author} {\bibfnamefont {H.}~\bibnamefont {Walther}},\ }\bibfield
  {title} {\bibinfo {title} {Quantum electrodynamic shifts of {R}ydberg energy
  levels between parallel metal plates},\ }\href
  {https://doi.org/10.1103/PhysRevLett.81.5784} {\bibfield  {journal} {\bibinfo
   {journal} {Phys. Rev. Lett.}\ }\textbf {\bibinfo {volume} {81}},\ \bibinfo
  {pages} {5784} (\bibinfo {year} {1998})}\BibitemShut {NoStop}%
\bibitem [{\citenamefont {Rentrop}\ \emph {et~al.}(2016)\citenamefont
  {Rentrop}, \citenamefont {Trautmann}, \citenamefont {Olivares}, \citenamefont
  {Jendrzejewski}, \citenamefont {Komnik},\ and\ \citenamefont
  {Oberthaler}}]{Rentrop16}%
  \BibitemOpen
  \bibfield  {author} {\bibinfo {author} {\bibfnamefont {T.}~\bibnamefont
  {Rentrop}}, \bibinfo {author} {\bibfnamefont {A.}~\bibnamefont {Trautmann}},
  \bibinfo {author} {\bibfnamefont {F.~A.}\ \bibnamefont {Olivares}}, \bibinfo
  {author} {\bibfnamefont {F.}~\bibnamefont {Jendrzejewski}}, \bibinfo {author}
  {\bibfnamefont {A.}~\bibnamefont {Komnik}},\ and\ \bibinfo {author}
  {\bibfnamefont {M.~K.}\ \bibnamefont {Oberthaler}},\ }\bibfield  {title}
  {\bibinfo {title} {Observation of the phononic {L}amb shift with a synthetic
  vacuum},\ }\href {https://doi.org/10.1103/PhysRevX.6.041041} {\bibfield
  {journal} {\bibinfo  {journal} {Phys. Rev. X}\ }\textbf {\bibinfo {volume}
  {6}},\ \bibinfo {pages} {041041} (\bibinfo {year} {2016})}\BibitemShut
  {NoStop}%
\bibitem [{\citenamefont {Fragner}\ \emph {et~al.}(2008)\citenamefont
  {Fragner}, \citenamefont {G{\"o}ppl}, \citenamefont {Fink}, \citenamefont
  {Baur}, \citenamefont {Bianchetti}, \citenamefont {Leek}, \citenamefont
  {Blais},\ and\ \citenamefont {Wallraff}}]{Fragner08}%
  \BibitemOpen
  \bibfield  {author} {\bibinfo {author} {\bibfnamefont {A.}~\bibnamefont
  {Fragner}}, \bibinfo {author} {\bibfnamefont {M.}~\bibnamefont {G{\"o}ppl}},
  \bibinfo {author} {\bibfnamefont {J.}~\bibnamefont {Fink}}, \bibinfo {author}
  {\bibfnamefont {M.}~\bibnamefont {Baur}}, \bibinfo {author} {\bibfnamefont
  {R.}~\bibnamefont {Bianchetti}}, \bibinfo {author} {\bibfnamefont
  {P.}~\bibnamefont {Leek}}, \bibinfo {author} {\bibfnamefont {A.}~\bibnamefont
  {Blais}},\ and\ \bibinfo {author} {\bibfnamefont {A.}~\bibnamefont
  {Wallraff}},\ }\bibfield  {title} {\bibinfo {title} {Resolving vacuum
  fluctuations in an electrical circuit by measuring the {L}amb shift},\ }\href
  {https://doi.org/10.1126/science.1164482} {\bibfield  {journal} {\bibinfo
  {journal} {Science}\ }\textbf {\bibinfo {volume} {322}},\ \bibinfo {pages}
  {1357} (\bibinfo {year} {2008})}\BibitemShut {NoStop}%
\bibitem [{\citenamefont {Yoshihara}\ \emph {et~al.}(2018)\citenamefont
  {Yoshihara}, \citenamefont {Fuse}, \citenamefont {Ao}, \citenamefont
  {Ashhab}, \citenamefont {Kakuyanagi}, \citenamefont {Saito}, \citenamefont
  {Aoki}, \citenamefont {Koshino},\ and\ \citenamefont {Semba}}]{Yoshihara18}%
  \BibitemOpen
  \bibfield  {author} {\bibinfo {author} {\bibfnamefont {F.}~\bibnamefont
  {Yoshihara}}, \bibinfo {author} {\bibfnamefont {T.}~\bibnamefont {Fuse}},
  \bibinfo {author} {\bibfnamefont {Z.}~\bibnamefont {Ao}}, \bibinfo {author}
  {\bibfnamefont {S.}~\bibnamefont {Ashhab}}, \bibinfo {author} {\bibfnamefont
  {K.}~\bibnamefont {Kakuyanagi}}, \bibinfo {author} {\bibfnamefont
  {S.}~\bibnamefont {Saito}}, \bibinfo {author} {\bibfnamefont
  {T.}~\bibnamefont {Aoki}}, \bibinfo {author} {\bibfnamefont {K.}~\bibnamefont
  {Koshino}},\ and\ \bibinfo {author} {\bibfnamefont {K.}~\bibnamefont
  {Semba}},\ }\bibfield  {title} {\bibinfo {title} {Inversion of qubit energy
  levels in qubit-oscillator circuits in the deep-strong-coupling regime},\
  }\href {https://doi.org/10.1103/PhysRevLett.120.183601} {\bibfield  {journal}
  {\bibinfo  {journal} {Phys. Rev. Lett.}\ }\textbf {\bibinfo {volume} {120}},\
  \bibinfo {pages} {183601} (\bibinfo {year} {2018})}\BibitemShut {NoStop}%
\bibitem [{\citenamefont {Mirhosseini}\ \emph {et~al.}(2018)\citenamefont
  {Mirhosseini}, \citenamefont {Kim}, \citenamefont {Ferreira}, \citenamefont
  {Kalaee}, \citenamefont {Sipahigil}, \citenamefont {Keller},\ and\
  \citenamefont {Painter}}]{Mirhosseini18}%
  \BibitemOpen
  \bibfield  {author} {\bibinfo {author} {\bibfnamefont {M.}~\bibnamefont
  {Mirhosseini}}, \bibinfo {author} {\bibfnamefont {E.}~\bibnamefont {Kim}},
  \bibinfo {author} {\bibfnamefont {V.~S.}\ \bibnamefont {Ferreira}}, \bibinfo
  {author} {\bibfnamefont {M.}~\bibnamefont {Kalaee}}, \bibinfo {author}
  {\bibfnamefont {A.}~\bibnamefont {Sipahigil}}, \bibinfo {author}
  {\bibfnamefont {A.~J.}\ \bibnamefont {Keller}},\ and\ \bibinfo {author}
  {\bibfnamefont {O.}~\bibnamefont {Painter}},\ }\bibfield  {title} {\bibinfo
  {title} {Superconducting metamaterials for waveguide quantum
  electrodynamics},\ }\href {https://doi.org/10.1038/s41467-018-06142-z}
  {\bibfield  {journal} {\bibinfo  {journal} {Nat. Commun.}\ }\textbf {\bibinfo
  {volume} {9}},\ \bibinfo {pages} {3706} (\bibinfo {year} {2018})}\BibitemShut
  {NoStop}%
\bibitem [{\citenamefont {Silveri}\ \emph {et~al.}(2019)\citenamefont
  {Silveri}, \citenamefont {Masuda}, \citenamefont {Sevriuk}, \citenamefont
  {Tan}, \citenamefont {Jenei}, \citenamefont {Hyypp{\"a}}, \citenamefont
  {Hassler}, \citenamefont {Partanen}, \citenamefont {Goetz},\ and\
  \citenamefont {{Lake et al.}}}]{Silveri19}%
  \BibitemOpen
  \bibfield  {author} {\bibinfo {author} {\bibfnamefont {M.}~\bibnamefont
  {Silveri}}, \bibinfo {author} {\bibfnamefont {S.}~\bibnamefont {Masuda}},
  \bibinfo {author} {\bibfnamefont {V.}~\bibnamefont {Sevriuk}}, \bibinfo
  {author} {\bibfnamefont {K.~Y.}\ \bibnamefont {Tan}}, \bibinfo {author}
  {\bibfnamefont {M.}~\bibnamefont {Jenei}}, \bibinfo {author} {\bibfnamefont
  {E.}~\bibnamefont {Hyypp{\"a}}}, \bibinfo {author} {\bibfnamefont
  {F.}~\bibnamefont {Hassler}}, \bibinfo {author} {\bibfnamefont
  {M.}~\bibnamefont {Partanen}}, \bibinfo {author} {\bibfnamefont
  {J.}~\bibnamefont {Goetz}},\ and\ \bibinfo {author} {\bibfnamefont {R.~E.}\
  \bibnamefont {{Lake et al.}}},\ }\bibfield  {title} {\bibinfo {title}
  {Broadband {L}amb shift in an engineered quantum system},\ }\href
  {https://doi.org/10.1038/s41567-019-0449-0} {\bibfield  {journal} {\bibinfo
  {journal} {Nat. Phys.}\ }\textbf {\bibinfo {volume} {15}},\ \bibinfo {pages}
  {533} (\bibinfo {year} {2019})}\BibitemShut {NoStop}%
\bibitem [{\citenamefont {Tan}\ \emph {et~al.}(2017)\citenamefont {Tan},
  \citenamefont {Partanen}, \citenamefont {Lake}, \citenamefont {Govenius},
  \citenamefont {Masuda},\ and\ \citenamefont {M{\"o}tt{\"o}nen}}]{Tan16}%
  \BibitemOpen
  \bibfield  {author} {\bibinfo {author} {\bibfnamefont {K.~Y.}\ \bibnamefont
  {Tan}}, \bibinfo {author} {\bibfnamefont {M.}~\bibnamefont {Partanen}},
  \bibinfo {author} {\bibfnamefont {R.~E.}\ \bibnamefont {Lake}}, \bibinfo
  {author} {\bibfnamefont {J.}~\bibnamefont {Govenius}}, \bibinfo {author}
  {\bibfnamefont {S.}~\bibnamefont {Masuda}},\ and\ \bibinfo {author}
  {\bibfnamefont {M.}~\bibnamefont {M{\"o}tt{\"o}nen}},\ }\bibfield  {title}
  {\bibinfo {title} {Quantum-circuit refrigerator},\ }\href
  {https://doi.org/10.1038/ncomms15189} {\bibfield  {journal} {\bibinfo
  {journal} {Nat. Commun.}\ }\textbf {\bibinfo {volume} {8}},\ \bibinfo {pages}
  {15189} (\bibinfo {year} {2017})}\BibitemShut {NoStop}%
\bibitem [{\citenamefont {Carmichael}(1999)}]{Carmichael99}%
  \BibitemOpen
  \bibfield  {author} {\bibinfo {author} {\bibfnamefont {H.~J.}\ \bibnamefont
  {Carmichael}},\ }\href {https://doi.org/10.1007/978-3-662-03875-8} {\emph
  {\bibinfo {title} {Statistical Methods in Quantum Optics~1}}}\ (\bibinfo
  {publisher} {Springer},\ \bibinfo {address} {Berlin},\ \bibinfo {year}
  {1999})\BibitemShut {NoStop}%
\bibitem [{\citenamefont {Wen}\ \emph {et~al.}(2019)\citenamefont {Wen},
  \citenamefont {Lin}, \citenamefont {Kockum}, \citenamefont {Suri},
  \citenamefont {Ian}, \citenamefont {Chen}, \citenamefont {Mao}, \citenamefont
  {Chiu}, \citenamefont {Delsing},\ and\ \citenamefont {{Nori et
  al.}}}]{Wen2019}%
  \BibitemOpen
  \bibfield  {author} {\bibinfo {author} {\bibfnamefont {P.~Y.}\ \bibnamefont
  {Wen}}, \bibinfo {author} {\bibfnamefont {K.-T.}\ \bibnamefont {Lin}},
  \bibinfo {author} {\bibfnamefont {A.~F.}\ \bibnamefont {Kockum}}, \bibinfo
  {author} {\bibfnamefont {B.}~\bibnamefont {Suri}}, \bibinfo {author}
  {\bibfnamefont {H.}~\bibnamefont {Ian}}, \bibinfo {author} {\bibfnamefont
  {J.~C.}\ \bibnamefont {Chen}}, \bibinfo {author} {\bibfnamefont {S.~Y.}\
  \bibnamefont {Mao}}, \bibinfo {author} {\bibfnamefont {C.~C.}\ \bibnamefont
  {Chiu}}, \bibinfo {author} {\bibfnamefont {P.}~\bibnamefont {Delsing}},\ and\
  \bibinfo {author} {\bibfnamefont {F.}~\bibnamefont {{Nori et al.}}},\
  }\bibfield  {title} {\bibinfo {title} {Large collective {L}amb shift of two
  distant superconducting artificial atoms},\ }\href
  {https://doi.org/10.1103/PhysRevLett.123.233602} {\bibfield  {journal}
  {\bibinfo  {journal} {Phys. Rev. Lett.}\ }\textbf {\bibinfo {volume} {123}},\
  \bibinfo {pages} {233602} (\bibinfo {year} {2019})}\BibitemShut {NoStop}%
\bibitem [{\citenamefont {Clerk}\ \emph {et~al.}(2020)\citenamefont {Clerk},
  \citenamefont {Lehnert}, \citenamefont {Bertet}, \citenamefont {Petta},\ and\
  \citenamefont {Nakamura}}]{Clerk20}%
  \BibitemOpen
  \bibfield  {author} {\bibinfo {author} {\bibfnamefont {A.~A.}\ \bibnamefont
  {Clerk}}, \bibinfo {author} {\bibfnamefont {K.~W.}\ \bibnamefont {Lehnert}},
  \bibinfo {author} {\bibfnamefont {P.}~\bibnamefont {Bertet}}, \bibinfo
  {author} {\bibfnamefont {J.~R.}\ \bibnamefont {Petta}},\ and\ \bibinfo
  {author} {\bibfnamefont {Y.}~\bibnamefont {Nakamura}},\ }\bibfield  {title}
  {\bibinfo {title} {Hybrid quantum systems with circuit quantum
  electrodynamics},\ }\href {https://doi.org/10.1038/s41567-020-0797-9}
  {\bibfield  {journal} {\bibinfo  {journal} {Nat. Phys.}\ }\textbf {\bibinfo
  {volume} {16}},\ \bibinfo {pages} {257} (\bibinfo {year} {2020})}\BibitemShut
  {NoStop}%
\bibitem [{\citenamefont {Lachance-Quirion}\ \emph {et~al.}(2020)\citenamefont
  {Lachance-Quirion}, \citenamefont {Wolski}, \citenamefont {Tabuchi},
  \citenamefont {Kono}, \citenamefont {Usami},\ and\ \citenamefont
  {Nakamura}}]{Lachance20}%
  \BibitemOpen
  \bibfield  {author} {\bibinfo {author} {\bibfnamefont {D.}~\bibnamefont
  {Lachance-Quirion}}, \bibinfo {author} {\bibfnamefont {S.~P.}\ \bibnamefont
  {Wolski}}, \bibinfo {author} {\bibfnamefont {Y.}~\bibnamefont {Tabuchi}},
  \bibinfo {author} {\bibfnamefont {S.}~\bibnamefont {Kono}}, \bibinfo {author}
  {\bibfnamefont {K.}~\bibnamefont {Usami}},\ and\ \bibinfo {author}
  {\bibfnamefont {Y.}~\bibnamefont {Nakamura}},\ }\bibfield  {title} {\bibinfo
  {title} {Entanglement-based single-shot detection of a single magnon with a
  superconducting qubit},\ }\href {https://doi.org/10.1126/science.aaz9236}
  {\bibfield  {journal} {\bibinfo  {journal} {Science}\ }\textbf {\bibinfo
  {volume} {367}},\ \bibinfo {pages} {425} (\bibinfo {year}
  {2020})}\BibitemShut {NoStop}%
\bibitem [{\citenamefont {Barzanjeh}\ \emph {et~al.}(2019)\citenamefont
  {Barzanjeh}, \citenamefont {Redchenko}, \citenamefont {Peruzzo},
  \citenamefont {Wulf}, \citenamefont {Lewis}, \citenamefont {Arnold},\ and\
  \citenamefont {Fink}}]{Barzanjeh19}%
  \BibitemOpen
  \bibfield  {author} {\bibinfo {author} {\bibfnamefont {S.}~\bibnamefont
  {Barzanjeh}}, \bibinfo {author} {\bibfnamefont {E.}~\bibnamefont
  {Redchenko}}, \bibinfo {author} {\bibfnamefont {M.}~\bibnamefont {Peruzzo}},
  \bibinfo {author} {\bibfnamefont {M.}~\bibnamefont {Wulf}}, \bibinfo {author}
  {\bibfnamefont {D.}~\bibnamefont {Lewis}}, \bibinfo {author} {\bibfnamefont
  {G.}~\bibnamefont {Arnold}},\ and\ \bibinfo {author} {\bibfnamefont {J.~M.}\
  \bibnamefont {Fink}},\ }\bibfield  {title} {\bibinfo {title} {Stationary
  entangled radiation from micromechanical motion},\ }\href
  {https://doi.org/10.1038/s41586-019-1320-2} {\bibfield  {journal} {\bibinfo
  {journal} {Nature}\ }\textbf {\bibinfo {volume} {570}},\ \bibinfo {pages}
  {480} (\bibinfo {year} {2019})}\BibitemShut {NoStop}%
\bibitem [{\citenamefont {Silveri}\ \emph {et~al.}(2017)\citenamefont
  {Silveri}, \citenamefont {Grabert}, \citenamefont {Masuda}, \citenamefont
  {Tan},\ and\ \citenamefont {M{\"o}tt{\"o}nen}}]{Silveri17}%
  \BibitemOpen
  \bibfield  {author} {\bibinfo {author} {\bibfnamefont {M.}~\bibnamefont
  {Silveri}}, \bibinfo {author} {\bibfnamefont {H.}~\bibnamefont {Grabert}},
  \bibinfo {author} {\bibfnamefont {S.}~\bibnamefont {Masuda}}, \bibinfo
  {author} {\bibfnamefont {K.~Y.}\ \bibnamefont {Tan}},\ and\ \bibinfo {author}
  {\bibfnamefont {M.}~\bibnamefont {M{\"o}tt{\"o}nen}},\ }\bibfield  {title}
  {\bibinfo {title} {Theory of quantum-circuit refrigeration by photon-assisted
  electron tunneling},\ }\href {https://doi.org/10.1103/PhysRevB.96.094524}
  {\bibfield  {journal} {\bibinfo  {journal} {Phys. Rev. B}\ }\textbf {\bibinfo
  {volume} {96}},\ \bibinfo {pages} {094524} (\bibinfo {year}
  {2017})}\BibitemShut {NoStop}%
\bibitem [{\citenamefont {Gely}\ \emph {et~al.}(2018)\citenamefont {Gely},
  \citenamefont {Steele},\ and\ \citenamefont {Bothner}}]{Gely18}%
  \BibitemOpen
  \bibfield  {author} {\bibinfo {author} {\bibfnamefont {M.~F.}\ \bibnamefont
  {Gely}}, \bibinfo {author} {\bibfnamefont {G.~A.}\ \bibnamefont {Steele}},\
  and\ \bibinfo {author} {\bibfnamefont {D.}~\bibnamefont {Bothner}},\
  }\bibfield  {title} {\bibinfo {title} {Nature of the {L}amb shift in weakly
  anharmonic atoms: {F}rom normal-mode splitting to quantum fluctuations},\
  }\href {https://doi.org/10.1103/PhysRevA.98.053808} {\bibfield  {journal}
  {\bibinfo  {journal} {Phys. Rev. A}\ }\textbf {\bibinfo {volume} {98}},\
  \bibinfo {pages} {053808} (\bibinfo {year} {2018})}\BibitemShut {NoStop}%
\bibitem [{\citenamefont {Sevriuk}\ \emph {et~al.}(2019)\citenamefont
  {Sevriuk}, \citenamefont {Tan}, \citenamefont {Hyypp{\"a}}, \citenamefont
  {Silveri}, \citenamefont {Partanen}, \citenamefont {Jenei}, \citenamefont
  {Masuda}, \citenamefont {Goetz}, \citenamefont {Vesterinen},\ and\
  \citenamefont {{Gr{\"o}nberg et al.}}}]{Sevriuk2019}%
  \BibitemOpen
  \bibfield  {author} {\bibinfo {author} {\bibfnamefont {V.}~\bibnamefont
  {Sevriuk}}, \bibinfo {author} {\bibfnamefont {K.~Y.}\ \bibnamefont {Tan}},
  \bibinfo {author} {\bibfnamefont {E.}~\bibnamefont {Hyypp{\"a}}}, \bibinfo
  {author} {\bibfnamefont {M.}~\bibnamefont {Silveri}}, \bibinfo {author}
  {\bibfnamefont {M.}~\bibnamefont {Partanen}}, \bibinfo {author}
  {\bibfnamefont {M.}~\bibnamefont {Jenei}}, \bibinfo {author} {\bibfnamefont
  {S.}~\bibnamefont {Masuda}}, \bibinfo {author} {\bibfnamefont
  {J.}~\bibnamefont {Goetz}}, \bibinfo {author} {\bibfnamefont
  {V.}~\bibnamefont {Vesterinen}},\ and\ \bibinfo {author} {\bibfnamefont
  {L.}~\bibnamefont {{Gr{\"o}nberg et al.}}},\ }\bibfield  {title} {\bibinfo
  {title} {Fast control of dissipation in a superconducting resonator},\ }\href
  {https://doi.org/10.1063/1.5116659} {\bibfield  {journal} {\bibinfo
  {journal} {Appl. Phys. Lett.}\ }\textbf {\bibinfo {volume} {115}},\ \bibinfo
  {pages} {082601} (\bibinfo {year} {2019})}\BibitemShut {NoStop}%
\bibitem [{\citenamefont {Hsu}\ \emph {et~al.}(2020)\citenamefont {Hsu},
  \citenamefont {Silveri}, \citenamefont {Gunyh\'o}, \citenamefont {Goetz},
  \citenamefont {Catelani},\ and\ \citenamefont {M\"ott\"onen}}]{Hsu20}%
  \BibitemOpen
  \bibfield  {author} {\bibinfo {author} {\bibfnamefont {H.}~\bibnamefont
  {Hsu}}, \bibinfo {author} {\bibfnamefont {M.}~\bibnamefont {Silveri}},
  \bibinfo {author} {\bibfnamefont {A.}~\bibnamefont {Gunyh\'o}}, \bibinfo
  {author} {\bibfnamefont {J.}~\bibnamefont {Goetz}}, \bibinfo {author}
  {\bibfnamefont {G.}~\bibnamefont {Catelani}},\ and\ \bibinfo {author}
  {\bibfnamefont {M.}~\bibnamefont {M\"ott\"onen}},\ }\bibfield  {title}
  {\bibinfo {title} {Tunable refrigerator for nonlinear quantum electric
  circuits},\ }\href {https://doi.org/10.1103/PhysRevB.101.235422} {\bibfield
  {journal} {\bibinfo  {journal} {Phys. Rev. B}\ }\textbf {\bibinfo {volume}
  {101}},\ \bibinfo {pages} {235422} (\bibinfo {year} {2020})}\BibitemShut
  {NoStop}%
\bibitem [{\citenamefont {Masuda}\ \emph {et~al.}(2018)\citenamefont {Masuda},
  \citenamefont {Tan}, \citenamefont {Partanen}, \citenamefont {Lake},
  \citenamefont {Govenius}, \citenamefont {Silveri}, \citenamefont {Grabert},\
  and\ \citenamefont {M{\"o}tt{\"o}nen}}]{Masuda2018}%
  \BibitemOpen
  \bibfield  {author} {\bibinfo {author} {\bibfnamefont {S.}~\bibnamefont
  {Masuda}}, \bibinfo {author} {\bibfnamefont {K.~Y.}\ \bibnamefont {Tan}},
  \bibinfo {author} {\bibfnamefont {M.}~\bibnamefont {Partanen}}, \bibinfo
  {author} {\bibfnamefont {R.~E.}\ \bibnamefont {Lake}}, \bibinfo {author}
  {\bibfnamefont {J.}~\bibnamefont {Govenius}}, \bibinfo {author}
  {\bibfnamefont {M.}~\bibnamefont {Silveri}}, \bibinfo {author} {\bibfnamefont
  {H.}~\bibnamefont {Grabert}},\ and\ \bibinfo {author} {\bibfnamefont
  {M.}~\bibnamefont {M{\"o}tt{\"o}nen}},\ }\bibfield  {title} {\bibinfo {title}
  {Observation of microwave absorption and emission from incoherent electron
  tunneling through a normal-metal-insulator-superconductor junction},\ }\href
  {https://doi.org/10.1038/s41598-018-21772-5} {\bibfield  {journal} {\bibinfo
  {journal} {Sci. Rep.}\ }\textbf {\bibinfo {volume} {8}},\ \bibinfo {pages}
  {3966} (\bibinfo {year} {2018})}\BibitemShut {NoStop}%
\bibitem [{\citenamefont {Hyypp\"a}\ \emph {et~al.}(2019)\citenamefont
  {Hyypp\"a}, \citenamefont {Jenei}, \citenamefont {Masuda}, \citenamefont
  {Sevriuk}, \citenamefont {Tan}, \citenamefont {Silveri}, \citenamefont
  {Goetz}, \citenamefont {Partanen}, \citenamefont {Lake},\ and\ \citenamefont
  {{Gr\"onberg et al.}}}]{Hyyppa19}%
  \BibitemOpen
  \bibfield  {author} {\bibinfo {author} {\bibfnamefont {E.}~\bibnamefont
  {Hyypp\"a}}, \bibinfo {author} {\bibfnamefont {M.}~\bibnamefont {Jenei}},
  \bibinfo {author} {\bibfnamefont {S.}~\bibnamefont {Masuda}}, \bibinfo
  {author} {\bibfnamefont {V.}~\bibnamefont {Sevriuk}}, \bibinfo {author}
  {\bibfnamefont {K.~Y.}\ \bibnamefont {Tan}}, \bibinfo {author} {\bibfnamefont
  {M.}~\bibnamefont {Silveri}}, \bibinfo {author} {\bibfnamefont
  {J.}~\bibnamefont {Goetz}}, \bibinfo {author} {\bibfnamefont
  {M.}~\bibnamefont {Partanen}}, \bibinfo {author} {\bibfnamefont {R.~E.}\
  \bibnamefont {Lake}},\ and\ \bibinfo {author} {\bibfnamefont
  {L.}~\bibnamefont {{Gr\"onberg et al.}}},\ }\bibfield  {title} {\bibinfo
  {title} {Calibration of cryogenic amplification chains using
  normal-metal–insulator–superconductor junctions},\ }\href
  {https://doi.org/10.1063/1.5096262} {\bibfield  {journal} {\bibinfo
  {journal} {Appl. Phys. Lett.}\ }\textbf {\bibinfo {volume} {114}},\ \bibinfo
  {pages} {192603} (\bibinfo {year} {2019})}\BibitemShut {NoStop}%
\bibitem [{\citenamefont {Fano}(1961)}]{Fano61}%
  \BibitemOpen
  \bibfield  {author} {\bibinfo {author} {\bibfnamefont {U.}~\bibnamefont
  {Fano}},\ }\bibfield  {title} {\bibinfo {title} {Effects of configuration
  interaction on intensities and phase shifts},\ }\href
  {https://doi.org/10.1103/PhysRev.124.1866} {\bibfield  {journal} {\bibinfo
  {journal} {Phys. Rev.}\ }\textbf {\bibinfo {volume} {124}},\ \bibinfo {pages}
  {1866} (\bibinfo {year} {1961})}\BibitemShut {NoStop}%
\bibitem [{\citenamefont {Wallraff}\ \emph {et~al.}(2004)\citenamefont
  {Wallraff}, \citenamefont {Schuster}, \citenamefont {Blais}, \citenamefont
  {Frunzio}, \citenamefont {Huang}, \citenamefont {Majer}, \citenamefont
  {Kumar}, \citenamefont {Girvin},\ and\ \citenamefont
  {Schoelkopf}}]{Wallraff04}%
  \BibitemOpen
  \bibfield  {author} {\bibinfo {author} {\bibfnamefont {A.}~\bibnamefont
  {Wallraff}}, \bibinfo {author} {\bibfnamefont {D.~I.}\ \bibnamefont
  {Schuster}}, \bibinfo {author} {\bibfnamefont {A.}~\bibnamefont {Blais}},
  \bibinfo {author} {\bibfnamefont {L.}~\bibnamefont {Frunzio}}, \bibinfo
  {author} {\bibfnamefont {R.-S.}\ \bibnamefont {Huang}}, \bibinfo {author}
  {\bibfnamefont {J.}~\bibnamefont {Majer}}, \bibinfo {author} {\bibfnamefont
  {S.}~\bibnamefont {Kumar}}, \bibinfo {author} {\bibfnamefont {S.~M.}\
  \bibnamefont {Girvin}},\ and\ \bibinfo {author} {\bibfnamefont {R.~J.}\
  \bibnamefont {Schoelkopf}},\ }\bibfield  {title} {\bibinfo {title} {Strong
  coupling of a single photon to a superconducting qubit using circuit quantum
  electrodynamics},\ }\href {https://doi.org/10.1038/nature02851} {\bibfield
  {journal} {\bibinfo  {journal} {Nature}\ }\textbf {\bibinfo {volume} {431}},\
  \bibinfo {pages} {162} (\bibinfo {year} {2004})}\BibitemShut {NoStop}%
\bibitem [{\citenamefont {Blais}\ \emph {et~al.}(2004)\citenamefont {Blais},
  \citenamefont {Huang}, \citenamefont {Wallraff}, \citenamefont {Girvin},\
  and\ \citenamefont {Schoelkopf}}]{Blais04}%
  \BibitemOpen
  \bibfield  {author} {\bibinfo {author} {\bibfnamefont {A.}~\bibnamefont
  {Blais}}, \bibinfo {author} {\bibfnamefont {R.-S.}\ \bibnamefont {Huang}},
  \bibinfo {author} {\bibfnamefont {A.}~\bibnamefont {Wallraff}}, \bibinfo
  {author} {\bibfnamefont {S.~M.}\ \bibnamefont {Girvin}},\ and\ \bibinfo
  {author} {\bibfnamefont {R.~J.}\ \bibnamefont {Schoelkopf}},\ }\bibfield
  {title} {\bibinfo {title} {Cavity quantum electrodynamics for superconducting
  electrical circuits: An architecture for quantum computation},\ }\href
  {https://doi.org/10.1103/PhysRevA.69.062320} {\bibfield  {journal} {\bibinfo
  {journal} {Phys. Rev. A}\ }\textbf {\bibinfo {volume} {69}},\ \bibinfo
  {pages} {062320} (\bibinfo {year} {2004})}\BibitemShut {NoStop}%
\bibitem [{\citenamefont {Abramowitz}\ and\ \citenamefont
  {Stegun}(1972)}]{Abramowitz72}%
  \BibitemOpen
  \bibfield  {author} {\bibinfo {author} {\bibfnamefont {M.}~\bibnamefont
  {Abramowitz}}\ and\ \bibinfo {author} {\bibfnamefont {I.~A.}\ \bibnamefont
  {Stegun}},\ }\href@noop {} {\emph {\bibinfo {title} {Handbook of Mathematical
  Functions}}}\ (\bibinfo  {publisher} {Dover},\ \bibinfo {address} {New
  York},\ \bibinfo {year} {1972})\BibitemShut {NoStop}%
\bibitem [{\citenamefont {Dynes}\ \emph {et~al.}(1978)\citenamefont {Dynes},
  \citenamefont {Narayanamurti},\ and\ \citenamefont {Garno}}]{Dynes78}%
  \BibitemOpen
  \bibfield  {author} {\bibinfo {author} {\bibfnamefont {R.~C.}\ \bibnamefont
  {Dynes}}, \bibinfo {author} {\bibfnamefont {V.}~\bibnamefont
  {Narayanamurti}},\ and\ \bibinfo {author} {\bibfnamefont {J.~P.}\
  \bibnamefont {Garno}},\ }\bibfield  {title} {\bibinfo {title} {Direct
  measurement of quasiparticle-lifetime broadening in a strong-coupled
  superconductor},\ }\href {https://doi.org/10.1103/PhysRevLett.41.1509}
  {\bibfield  {journal} {\bibinfo  {journal} {Phys. Rev. Lett.}\ }\textbf
  {\bibinfo {volume} {41}},\ \bibinfo {pages} {1509} (\bibinfo {year}
  {1978})}\BibitemShut {NoStop}%
\bibitem [{\citenamefont {Ingold}\ and\ \citenamefont
  {Nazarov}(1992)}]{Ingold92}%
  \BibitemOpen
  \bibfield  {author} {\bibinfo {author} {\bibfnamefont {G.-L.}\ \bibnamefont
  {Ingold}}\ and\ \bibinfo {author} {\bibfnamefont {Y.~V.}\ \bibnamefont
  {Nazarov}},\ }\bibfield  {title} {\bibinfo {title} {Charge tunneling rates in
  ultrasmall junctions},\ }in\ \href {http://arxiv.org/abs/cond-mat/0508728}
  {\emph {\bibinfo {booktitle} {Single Charge Tunneling: Coulomb Blockade
  Phenomena in Nanostructures}}}\ (\bibinfo  {publisher} {eds Grabert, H. \&
  Devoret, M.~H.) (Plenum},\ \bibinfo {address} {New York},\ \bibinfo {year}
  {1992})\BibitemShut {NoStop}%
\end{thebibliography}
\end{document}